\documentclass[english,aps,prc,twocolumn,amsmath,amssymb,preprintnumbers,
superscriptaddress,nofootinbib]{revtex4-1}
\usepackage[T1]{fontenc}
\usepackage[utf8]{inputenc}
\usepackage{babel}
\usepackage{amsmath}
\usepackage{graphicx}
\usepackage{color}
\usepackage{siunitx}
\usepackage{enumerate}
\usepackage{subcaption}
\PassOptionsToPackage{hyphens}{url}
\usepackage[unicode=true]{hyperref}
\usepackage[hyphens]{url}

\newcommand{\beq}{\begin{equation}}
\newcommand{\eeq}{\end{equation}}
\newcommand{\ba}{\begin{array}}
\newcommand{\ea}{\end{array}}
\newcommand{\bea}{\begin{eqnarray}}
\newcommand{\eea}{\end{eqnarray}}
\newcommand{\bi}{\begin{itemize}}  
\newcommand{\ei}{\end{itemize}}
\newcommand{\ben}{\begin{enumerate}} 
\newcommand{\een}{\end{enumerate}}
\newcommand{\bc}{\begin{center}}
\newcommand{\ec}{\end{center}}

\newcommand{\MeV}{{\rm MeV}}

\newcommand{\nsat}{n_{\text{sat}}}

\begin{document}

\title{Damping of density oscillations in neutrino-transparent nuclear matter}

\author{Mark G. Alford}
\affiliation{Physics Department, Washington University, St.~Louis, MO~63130, USA}
\author{Steven P. Harris}
\affiliation{Physics Department, Washington University, St.~Louis, MO~63130, USA}

\begin{abstract}
We calculate the bulk-viscous dissipation time for adiabatic density oscillations in
nuclear matter at densities of 1--7 times nuclear saturation density and at temperatures ranging from 1 MeV, where corrections to previous low-temperature calculations become important, up to 10 MeV, where the assumption of neutrino transparency is no longer valid. 
Under these conditions, which are expected to occur in neutron star mergers, damping of density oscillations arises from beta equilibration via weak interactions.  We find that for 1\,kHz oscillations the shortest dissipation times are in the 5 to 20\,ms range, depending on the equation of state, which means that bulk viscous damping could affect the dynamics of a neutron star merger. 
For higher frequencies the dissipation time can be even shorter.

\end{abstract}

\date{27 September 2019} 

\maketitle

\section{Introduction}
The observation of gravitational waves from neutron star mergers \cite{TheLIGOScientific:2017qsa}
has drawn attention to the importance of understanding the properties of nuclear matter at the densities and temperatures attained in mergers. Recent studies \cite{Abbott:2018exr,Landry:2018prl} of the gravitational wave signals estimate that the two neutron stars involved in GW170817 reached central densities of at least twice nuclear saturation density ($\nsat \equiv 0.16 \text{ fm}^{-3}$).  Numerical simulations \cite{Perego:2019adq,Hanauske:2019qgs,Hanauske:2017oxo,Baiotti:2016qnr,Kastaun:2016elu,Bernuzzi:2015opx,Foucart:2015gaa,kiuchi:2012mk,sekiguchi:2011zd} find that during the first 10 to 20 ms after the neutron stars make contact the material at densities of up to several times $\nsat$ can reach temperatures of many tens of MeV, perhaps up to 80 \cite{Hanauske:2019qgs,Hanauske:2017oxo} or even 100 MeV \cite{Perego:2019adq}, and co-moving fluid elements are subjected to strong density oscillations at a typical frequency of 1\,kHz \cite{Alford:2017rxf}. This raises the possibility of bulk viscosity in nuclear matter playing an important role if it is strong enough to damp those oscillations on a short enough timescale to affect the dynamics of the merger\footnote{We use the term ``merger'' to refer to the late stages of the inspiral as well as the process that begins when the stars touch.}.

An initial estimate for $npe\mu$ matter \cite{Alford:2017rxf} suggested that the bulk viscous dissipation time could be extremely short: in the range of a few milliseconds.  In this work we focus on $npe$ matter
in the neutrino-transparent regime ($T\lesssim 5$ to 10\,MeV \cite{Alford:2018lhf,Roberts:2016mwj,Haensel:1987zz,1979ApJ...230..859S,Sawyer:1975js}), where neutrinos escape from the merger region, and make a detailed study.  We calculate the dependence of the bulk viscous dissipation time on density and temperature, and we explore its sensitivity to the equation of state. We find that the lowest dissipation times occur at temperatures of about 3\,MeV, which lies in the neutrino-transparent regime, since neutrinos with energies of 3\,MeV have mean free paths on the order of a few kilometers\cite{Alford:2018lhf,Sawyer:1975js,1979ApJ...230..859S}. 
For very low densities ($n_B \approx 0.5\nsat$), the dissipation time of 1 kHz oscillations is as low as 5 ms.  At densities above $\nsat$, the dissipation times can be as low as 20 ms.  These times are similar for both equations of state that we study.

Bulk viscosity for oscillations in the kHz range arises from beta equilibration via weak interactions. Neutrino transparency means there is no Fermi sea of neutrinos, so in the relevant Urca processes neutrinos only occur in final states \cite{Yakovlev:2000jp,Shapiro:1983du}. This deviation from detailed balance leads to corrections to the standard Fermi Surface (FS) approximation when the temperature rises above about 1\,MeV \cite{Alford:2018lhf}. Those corrections are included in our calculations.

We work in natural units, where $\hbar=c=k_B=1$.  All data presented in our figures can be found in the Supplemental Material \cite{supp}.

\section{Nuclear matter and the Urca process}
\label{sec:nuclear}

Beta equilibration establishes the stable proton fraction
via Urca processes.  In this work we will focus on processes involving electrons, leaving muon contributions for future investigation.
If there is a deficit of protons, protons
are created via the neutron decay processes
\beq
\ba{rcl@{\quad}l}
n &\rightarrow & p + e^- + \bar{\nu}_e  &\mbox{direct Urca}\\
n+N &\rightarrow & N + p + e^- + \bar{\nu}_e &\mbox{modified Urca}
\ea
\eeq
Here, $N$ denotes a spectator neutron or proton.  
If there is a deficit of neutrons, neutrons are created
via the electron capture processes
\beq
\ba{rcl@{\quad}l}
e^- + p &\rightarrow& n + \nu_e & \mbox{direct Urca} \\
N + e^- + p &\rightarrow& N + n + \nu_e & \mbox{modified Urca}
\ea
\eeq

The direct Urca process is in general faster than the modified Urca process, since it involves two fewer particles.  

The rate of the direct Urca neutron decay and electron capture processes are given by the twelve dimensional phase space integrals \cite{Yakovlev:2000jp}
\begin{align}
\Gamma_{dU,nd} &= \int \dfrac{\mathop{d^3p_n}}{\left(2\pi\right)^3}\dfrac{\mathop{d^3p_p}}{\left(2\pi\right)^3}\dfrac{\mathop{d^3p_e}}{\left(2\pi\right)^3}\dfrac{\mathop{d^3p_{\nu}}}{\left(2\pi\right)^3}\frac{\sum_{\text{spins}}\vert\mathcal{M_{\text{dU}}}\vert^2}{2^4 E_n^*E_p^*E_eE_{\nu}} \label{eq:ndecay}\\
&\times\left(2\pi\right)^4\delta^4(p_n-p_p-p_e-p_{\nu}) f_n\left(1-f_p\right)\left(1-f_e\right)\nonumber \\[2ex]
\Gamma_{dU,ec} &= \int \dfrac{\mathop{d^3p_n}}{\left(2\pi\right)^3}\dfrac{\mathop{d^3p_p}}{\left(2\pi\right)^3}\dfrac{\mathop{d^3p_e}}{\left(2\pi\right)^3}\dfrac{\mathop{d^3p_{\nu}}}{\left(2\pi\right)^3}\frac{\sum_{\text{spins}}\vert\mathcal{M_{\text{dU}}}\vert^2}{2^4 E_n^*E_p^*E_eE_{\nu}} \label{eq:ecapture}\\
&\times\left(2\pi\right)^4\delta^4(p_n-p_p-p_e+p_{\nu}) \left(1-f_n\right)f_pf_e\nonumber,
\end{align}
where $f_i=1/\{1 + \exp[(E_i-\mu_i)/T]\}$ are the Fermi-Dirac distributions for species $i=n,p,\text{ or }e$ with chemical potential $\mu_i$ and the matrix element is \cite{Yakovlev:2000jp}
\begin{align}
\label{eq:durca_M}
\sum_{\text{spins}}\vert\mathcal{M_{\text{dU}}}\vert^2 &= 32G^2E_n^*E_p^*E_eE_{\nu} \nonumber\\
&\times \left(1+3g_A^2+\left(1-g_A^2\right)\dfrac{\mathbf{p}_e\cdot\mathbf{p}_{\nu}}{E_eE_{\nu}}\right),
\end{align}
where $G^2 = G_F^2\cos^2{\theta_c} = 1.29\times10^{-22}\MeV^{-4}$ and $g_A = 1.26$.  We will describe the nucleon energy dispersion relations and discuss the significance of using $E^*$ instead of $E$ for the nucleons in Sec.~\ref{sec:models}.  We note here that the matrix elements in \cite{Alford:2018lhf} denoted $\langle \vert \mathcal{M}\vert^2\rangle$ are in fact $\sum_{\text{spins}} \vert \mathcal{M}\vert^2 / (2E_1\times 2E_2 \times ... \times 2E_n)$ for a process with $n$ particles in the initial and final states combined.  

The rate of modified Urca neutron decay and electron capture with a nucleon spectator is given by the eighteen dimensional phase space integrals given in \cite{Haensel:2001mw}.   

\section{Bulk viscosity in the Fermi Surface approximation}
\subsection{Urca rates}

At low temperatures $T\lesssim 1\,\MeV$ \cite{Alford:2018lhf}, where the Fermi surfaces are sharply defined, we can make the approximation that only particles near the Fermi surface can participate in Urca processes.  We call this the ``Fermi Surface (FS) approximation''. At these temperatures the beta equilibrium condition is \cite{Yakovlev:2000jp,Shapiro:1983du}
\begin{equation}
\mu_n = \mu_p + \mu_e \quad \text{(Fermi Surface approx)}  \ . 
\label{eq:betaFS}
\end{equation}
This condition enforces the equality of Urca rates for proton production and proton capture, yielding a proton fraction that is constant in time \cite{Yuan:2005tj}.  In the Fermi Surface approximation the phase space integrals (Sec.~\ref{sec:nuclear}) can be simplified by fixing the momentum magnitudes to the corresponding Fermi momenta.

In the Fermi surface approximation, the direct Urca matrix element (Eq.~\ref{eq:durca_M}) simplifies to 
\begin{equation}
\sum_{\text{spins}} \vert \mathcal{M_{\text{dU}}}\vert^2 = 32G^2(1+3g_A^2)E_n^*E_p^*E_eE_{\nu}
\label{eq:mdUapprox}
\end{equation}
under the assumption of non-relativistic nucleons \cite{Yakovlev:2000jp}.
In the Fermi Surface approximation, in (low-temperature) beta-equilibrium (Eq.~\ref{eq:betaFS}) the rate of direct Urca neutron decay and direct Urca electron capture are equal and are given by \cite{Alford:2018lhf,Yakovlev:2000jp,Haensel:2000vz}
\begin{align}
\Gamma_{dU,nd} &= \Gamma_{dU,ec} =A_{dU} G^2\left(1+3g_A^2\right) m_nm_pp_{Fe}\vartheta_{dU}T^5\label{eq:dU_FS}\\
\vartheta_{dU} &\equiv \left\{\ba{ll}0 & \text{if}\ p_{Fn}>p_{Fp}+p_{Fe}\\
1 & \text{if}\ p_{Fn}<p_{Fp}+p_{Fe},
\ea \right. \nonumber \\[2ex]
A_{dU} &\equiv 3\left(\pi^2\zeta(3)+15\zeta(5)\right)/(16\pi^5)\approx 0.0170\nonumber \ .  
\end{align}
We see that in this approximation, the direct Urca process has a density threshold, above which the process is kinematically allowed.

In the Fermi surface approximation, the neutron-spectator modified Urca rates (both neutron decay and electron capture, which are equal in beta equilibrium (Eq.~\ref{eq:betaFS})) are given by
 \begin{align}
& \Gamma_{\text{mU,n}}
= A_{mU}G^2f_{\pi NN}^4g_A^2\dfrac{m_n^3m_p}{m_{\pi}^4} \dfrac{p_{Fn}^4p_{Fp}}{\left(p_{Fn}^2+m_{\pi}^2\right)^2} \, \vartheta_n \, T^7 \ , \label{eq:mU_n_FS}\\[2ex]
\vartheta_n &\equiv \left\{
\ba{ll}
1 & \text{if}\ p_{Fn}>p_{Fp}+p_{Fe}\\
1-\dfrac{3}{8}\dfrac{(p_{Fp}+p_{Fe}-p_{Fn})^2}{p_{Fp}p_{Fe}} & \text{if}\ p_{Fn}<p_{Fp}+p_{Fe}
\ea \right. \nonumber
\end{align}
and the proton-spectator modified Urca rates are given by \cite{Alford:2018lhf,Yakovlev:2000jp}
\begin{align}
\Gamma_{\text{mU,p}} &= \dfrac{A_{mU}}{7}G^2f_{\pi NN}^4g_A^2 \dfrac{m_nm_p^3}{m_{\pi}^4} \nonumber \\
&\times \dfrac{p_{Fn} (p_{Fn}\!-\!p_{Fp})^4}{\bigl((p_{Fn}\!-\!p_{Fp})^2+m_{\pi}^2\bigr)^2}\, \vartheta_p \, T^7 \label{eq:mU_p_FS}
\end{align}
\begin{align}
\vartheta_p &\equiv \left\{
\ba{ll}
\qquad 0 &\text{if} \ p_{Fn}>3p_{Fp}+p_{Fe}\\[3ex]
\dfrac{(3p_{Fp}+p_{Fe}-p_{Fn})^2}{p_{Fn}p_{Fe}} & \text{if}\ \ba{l} p_{Fn}>3p_{Fp}-p_{Fe}\\p_{Fn}<3p_{Fp}+p_{Fe}\ea \\[3ex]
4\dfrac{3p_{Fp}-p_{Fn}}{p_{Fn}}
& \text{if}\ \ba{l} 3 p_{Fp}-p_{Fe}>p_{Fn} \\ p_{Fn} >p_{Fp}+p_{Fe}\ea \\[3ex]
\Bigl( 2+3\dfrac{2p_{Fp}-p_{Fn}}{p_{Fe}} \\
\quad -\,3\dfrac{(p_{Fp}-p_{Fe})^2}{p_{Fn}p_{Fe}}\Bigr) & \text{if}\ p_{Fn}<p_{Fp}+p_{Fe} \ .
\ea\right. \nonumber
\end{align}

\subsection{Bulk viscosity}
\label{sec:bulk-visc-basics}

Consider a fluid element of nuclear matter subjected to a small-amplitude, periodic baryon density oscillation
\begin{equation}
    n_B(t) = n_B + \Delta n \sin{\omega t} \ ,
\end{equation}
where $\Delta n \ll n_B$.
Since the equilibrium proton fraction varies with density, changing the density will temporarily push the nuclear matter out of beta equilibrium
 by an amount  \cite{Yakovlev:2000jp,1992A&A...262..131H,fi09000a}
 \beq
 \mu_{\Delta}\equiv \mu_n-\mu_p-\mu_e \ .
 \eeq
 We will consider only ``subthermal'' density oscillations where $\mu_{\Delta} \ll T$.  In response to the density change, the nuclear matter will try to reestablish beta equilibrium via the Urca process which has a characteristic rate $\gamma(n_B,T)$ \cite{Yakovlev:2000jp}.  
The bulk viscosity of neutrino-transparent nuclear matter is given by \cite{Alford:2010gw,Schmitt:2017efp,Haensel:1992zz}
\begin{equation}
    \zeta = \frac{C^2}{B}\frac{\gamma}{\omega^2+\gamma^2}, 
    \label{eq:bv}
\end{equation}
where $B$ and $C$ are susceptibilities of the nuclear matter which depend on the equation of state but not on the weak-interaction equilibration rate,
\begin{align}
 B &= -\frac{1}{n_B} \frac{\partial\mu_{\Delta}}{\partial x_p} \bigg\rvert_{n_B} \ , \nonumber \\
 C &= n_B \frac{\partial \mu_{\Delta}}{\partial n_B}\bigg\rvert_{x_p} \ .
 \label{eq:B_C_defn}
\end{align}
In this paper we assume that there is negligible heat flow between adjacent fluid elements during the merger. This is valid as long as the thermal equilibration time in the absence of neutrinos is much longer than about 10\,ms. From Eq.~(1) of Ref.~\cite{Alford:2017rxf} this will be true as long as density oscillations (and the resultant thermal gradients) have wavelengths longer than about a meter. This criterion is obeyed in current simulations, whose spatial resolution is tens of meters at best.  Since heat does not flow we use adiabatic susceptibilities, evaluating the derivatives Eq.~\ref{eq:B_C_defn} at constant entropy per baryon $S/N_B$, or equivalently, at constant entropy density per baryon density $s/n_B$.  See App.~\ref{app:adiabatic} for a comparison of adiabatic and isothermal thermodynamic quantities.

The equilibration rate $\gamma = B\lambda$, where
\begin{equation}
    \lambda = \frac{\partial (\Gamma_{n\rightarrow p}-\Gamma_{p\rightarrow n})}{\partial \mu_{\Delta}}\bigg\rvert_{\mu_{\Delta}=0}.
\end{equation}
In the Fermi Surface approximation, we can analytically compute $\lambda$ from Eqs. \ref{eq:dU_FS},\ref{eq:mU_n_FS}, and \ref{eq:mU_p_FS},  \cite{Yakovlev:2000jp,Haensel:1992zz,Reisenegger:1994be,Friman:1978zq,Sawyer:1989dp}
\begin{equation}
    \lambda = \lambda_{dU}+\lambda_{mU\!,n}+\lambda_{mU\!,p},
    \label{eq:lambda}
\end{equation}
where
\begin{align}
    \lambda_{dU} &= \frac{17}{240\pi}G^2(1+3g_A^2)m_nm_pp_{Fe}T^4\ \ ,\\
    \lambda_{mU\!,n} &= \frac{367}{1152\pi^3}G^2g_A^2f_{\pi NN}^4 \frac{m_n^3 m_p}{m_{\pi}^4} \nonumber\\
    &\times \frac{p_{Fn}^4 p_{Fp}}{(p_{Fn}^2+m_{\pi}^2)^2}\vartheta_n T^6 \ ,\\
    \lambda_{mU\!,p}&= \frac{367}{8064\pi^3}G^2 g_A^2 f_{\pi NN}^4 \frac{m_n m_p^3}{m_{\pi}^4} \nonumber\\
    &\times \frac{p_{Fn}(p_{Fn}-p_{Fp})^4}{((p_{Fn}-p_{Fp})^2+m_{\pi}^2)^2}\vartheta_p T^6 \ .
\end{align}
From Eq.~\ref{eq:bv} it follows that, for an oscillation of fixed frequency $\omega$, the bulk viscosity has a resonant maximum when the equilibration rate (which varies as a function of density and temperature) coincides with the oscillation frequency, i.e., when $\gamma(n_B,T) = \omega$.  
A major goal of this paper will be to map the regions in density and temperature where this maximum is achieved.

\section{Intermediate temperatures}

\subsection{Urca process at intermediate temperatures}

Below the direct Urca threshold, neutrons, protons, and electrons on their Fermi surfaces cannot participate in the direct Urca process while conserving energy and momentum.  Direct Urca still occurs, but it involves particles away from their Fermi surfaces, and so the rate is Boltzmann suppressed.  As temperatures rise above 1 MeV, up to the neutrino trapping temperature, the Boltzmann suppression lessens, and the direct Urca rate becomes comparable to and then, eventually, greater than the modified Urca rate, thus broadening the direct Urca threshold.  In contrast, the presence of a spectator nucleon means that neutrons, protons, and electrons close to their Fermi surfaces can participate in a modified Urca process, and so the Fermi surface approximation is still appropriate for modified Urca, even at the moderately high temperatures that we consider \cite{Alford:2018lhf}.

The beta equilibrium condition becomes
\begin{equation}
    \mu_n = \mu_p + \mu_e + \mu_{\delta} \ ,
    \label{eq:full_beq}
\end{equation}
where the additional chemical potential $\mu_{\delta}$ arises from the absence of detailed balance: neutron decay and electron capture processes are not exact inverses of each other. $\mu_\delta$ is a function of temperature and baryon density, and its value is determined by the requirement that the rates for neutron decay and electron capture rates must balance \cite{Alford:2018lhf}. These rates include direct Urca contributions, calculated by integration of the full phase space, and modified Urca contributions for which we can use the Fermi Surface approximation as described in the appendix of Ref.~\cite{Alford:2018lhf}.

\begin{figure}
\includegraphics[width=.5\textwidth]{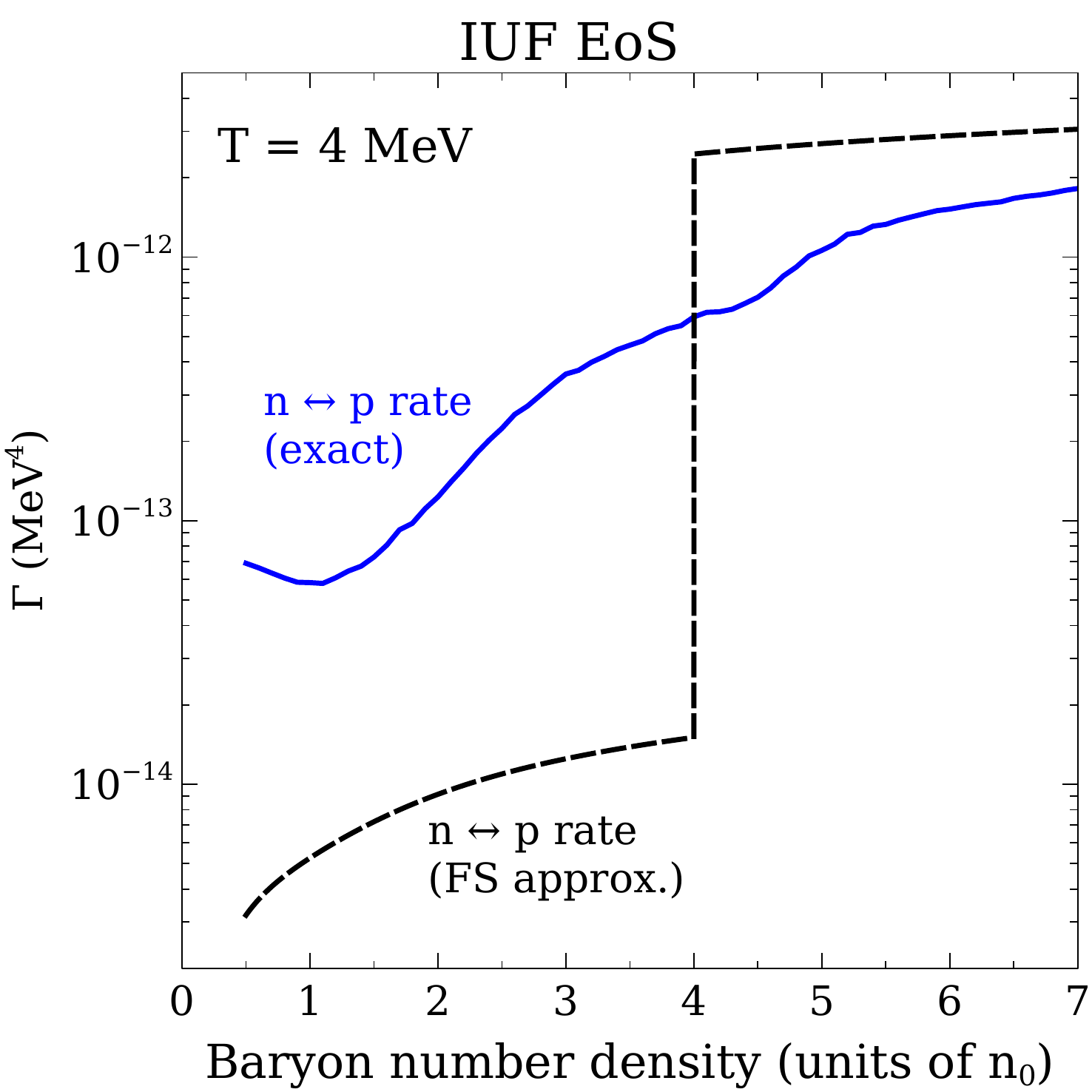}
\caption{Total Urca rates (direct plus modified) in beta equilibrium for the IUF EoS at $T=4 \text{ MeV}$.  The dashed (black) curve is the Fermi Surface approximation to the Urca rates, using the low-temperature beta equilibrium criterion Eq.~\ref{eq:betaFS}. The solid (blue) curve is the total Urca rate with the full phase space integral, and using the general beta equilibrium condition Eq.~\ref{eq:full_beq}}.
\label{fig:urca_rates}
\end{figure}

To illustrate the breakdown of the Fermi Surface approximation at temperatures relevant for
neutron star mergers, we show in
Fig.~\ref{fig:urca_rates} the comparison between rates calculated in the Fermi Surface approximation (using the low-temperature beta equilibrium condition Eq.~\ref{eq:betaFS}) and the full phase space integration (with the general beta equilibrium condition Eq.~\ref{eq:full_beq}). 
We show the neutron decay rate, which is equal to the electron capture rate in each case, calculated for neutrino-transparent nuclear matter described by the IUF equation of state, described below. We see that at $T=4\,\MeV$ the Fermi surface approximation makes the direct Urca threshold seem unphysically sharp, underestimating the below-threshold rates by an order of magnitude and overestimating the above-threshold rates by a factor of 3 to 5.
\subsection{Bulk viscosity at intermediate temperatures}
To calculate bulk viscosity at temperatures where the Fermi Surface approximation is not valid,
we use the appropriate characterization of the deviation from equilibrium,
\begin{equation}
    \mu_{\Delta} = \mu_n - \mu_p - \mu_e - \mu_{\delta} \ .
\end{equation}
When $\mu_\Delta=0$ the system is in true beta equilibrium (\ref{eq:full_beq}).  In this regime, the equilibration rate $\lambda$ (Eq.~\ref{eq:lambda}) no longer has a simple analytic form, as the direct Urca rates (\ref{eq:ndecay}) and (\ref{eq:ecapture}) can only be simplified to three-dimensional numerical integrals \cite{Alford:2018lhf}.  We obtain $\lambda$ by calculating the difference $\Gamma_{n\rightarrow p} - \Gamma_{p\rightarrow n}$  at proton fractions around the beta equilibrium value where the rates are equal.  The slope of the difference as a function of $\mu_{\Delta}$ at $\mu_{\Delta}=0$ is $\lambda$.

\section{Results} 
\subsection{Models of nuclear matter}
\label{sec:models}
To gauge the sensitivity of our results to the equation of state of nuclear matter, we will use two representative equations of state, one stiffer (DD2\cite{Hempel:2009mc,Typel:2009sy}) and one softer (IUF \cite{Hempel:2009mc,Fattoyev:2010mx,RocaMaza:2008ja}).  Both are tabulated at the CompOSE website (\url{https://compose.obspm.fr/eos/18/} and \url{https://compose.obspm.fr/eos/22/}).  Both are relativistic mean field theories, where nucleons interact strongly by exchanging sigma, omega, and rho mesons.  The couplings between the mesons and nuclei are chosen to reproduce nuclear observables like the nuclear binding energy, saturation density, symmetry energy, incompressibility, among others \cite{glendenning2000compact,Dutra:2014qga,Oertel:2016bki}.  Aside from the standard nucleon-meson linear couplings, the IUF model has self-interactions in the sigma and omega fields, plus interactions among the mesons (see Eq.~7 in \cite{Dutra:2014qga}).  DD2 has the standard linear nucleon-meson couplings, but instead of nonlinear self-interactions or meson-meson interactions, it promotes the nucleon-meson couplings to density-dependent functions \cite{Dutra:2014qga}.  DD2 has no direct Urca threshold (ie, direct Urca is kinematically forbidden at all densities in the Fermi surface approximation) (see Fig.~2 in \cite{Alford:2017rxf}), while IUF has a direct Urca threshold near $4n_{\text{sat}}$.  The maximum mass neutron star for DD2 is $2.42\,M_\odot$ and for IUF it is $1.96\,M_\odot$.

In our calculations we assume the neutrinos and electrons are ultrarelativistic free particles, and nucleonic excitations have the dispersion relation \cite{Roberts:2016mwj,Roberts:2012um}
\begin{equation}
    E_i = U_i + m_i + \frac{p_i^2}{2m_i} \quad \mbox{for}~i=(p,n)
\end{equation}
where $U_i$, the nuclear mean field, is chosen as a function of density and temperature so that the Fermi energy $E_{F,i}\equiv E_i(p_{F,i})$ matches the chemical potential $\mu_i$ for the given EoS.  For the mass $m_i$, we use the rest mass in vacuum.  The microscopic origin of the nuclear mean field can be understood through the framework of relativistic mean field theories, where the mean field $U$ is a function of the vacuum expectation values of the $\omega$ and $\rho$ mesons which are the strong force carriers between nucleons \cite{Fu:2008zzg}.

As alluded to in Sec.~\ref{sec:nuclear}, the neutron and proton energies can be written as $E_i = U_i + E_i^*$, where $E_i^* = m_i + p^{2}/(2m_i)$.  In the rate calculations (\ref{eq:ndecay}) and (\ref{eq:ecapture}), $E^*$ should be used for the energies in the matrix element and in the energy factors in the denominator, while $E$ should be used in the energy delta function and the Fermi Dirac factors \cite{Leinson:2002bw,Roberts:2016mwj,Fu:2008zzg}.  However, in the approximation we used for the direct Urca matrix element (\ref{eq:mdUapprox}), the $E^*$ factors cancel out.

\subsection{Bulk viscosity}
\label{sec:bv_results}
\begin{figure*}[t!]
    \centering
    \begin{subfigure}[t]{0.5\textwidth}
        \centering
        \includegraphics[width=.95\textwidth]{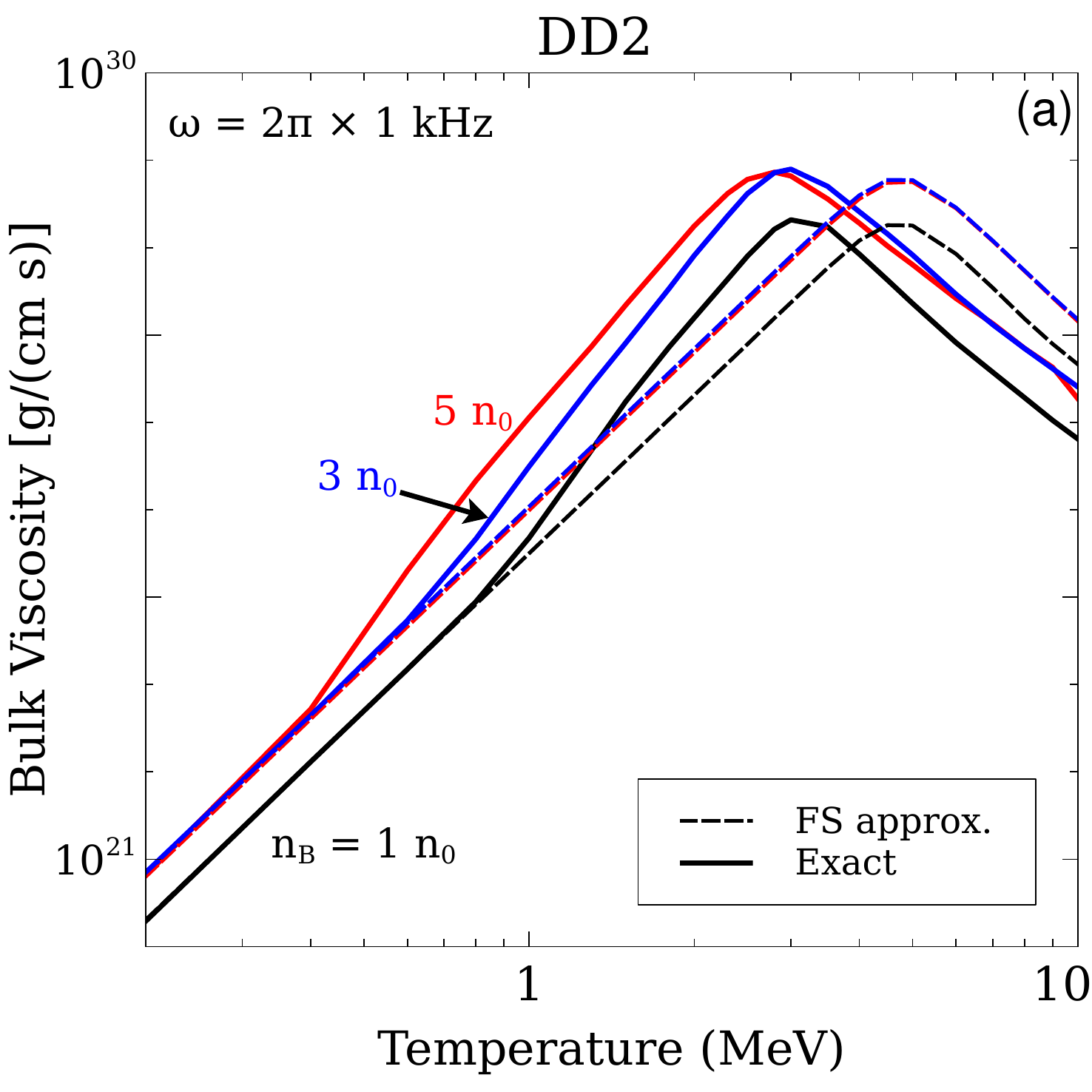}
    \end{subfigure}%
    \begin{subfigure}[t]{0.5\textwidth}
        \centering
        \includegraphics[width=.95\textwidth]{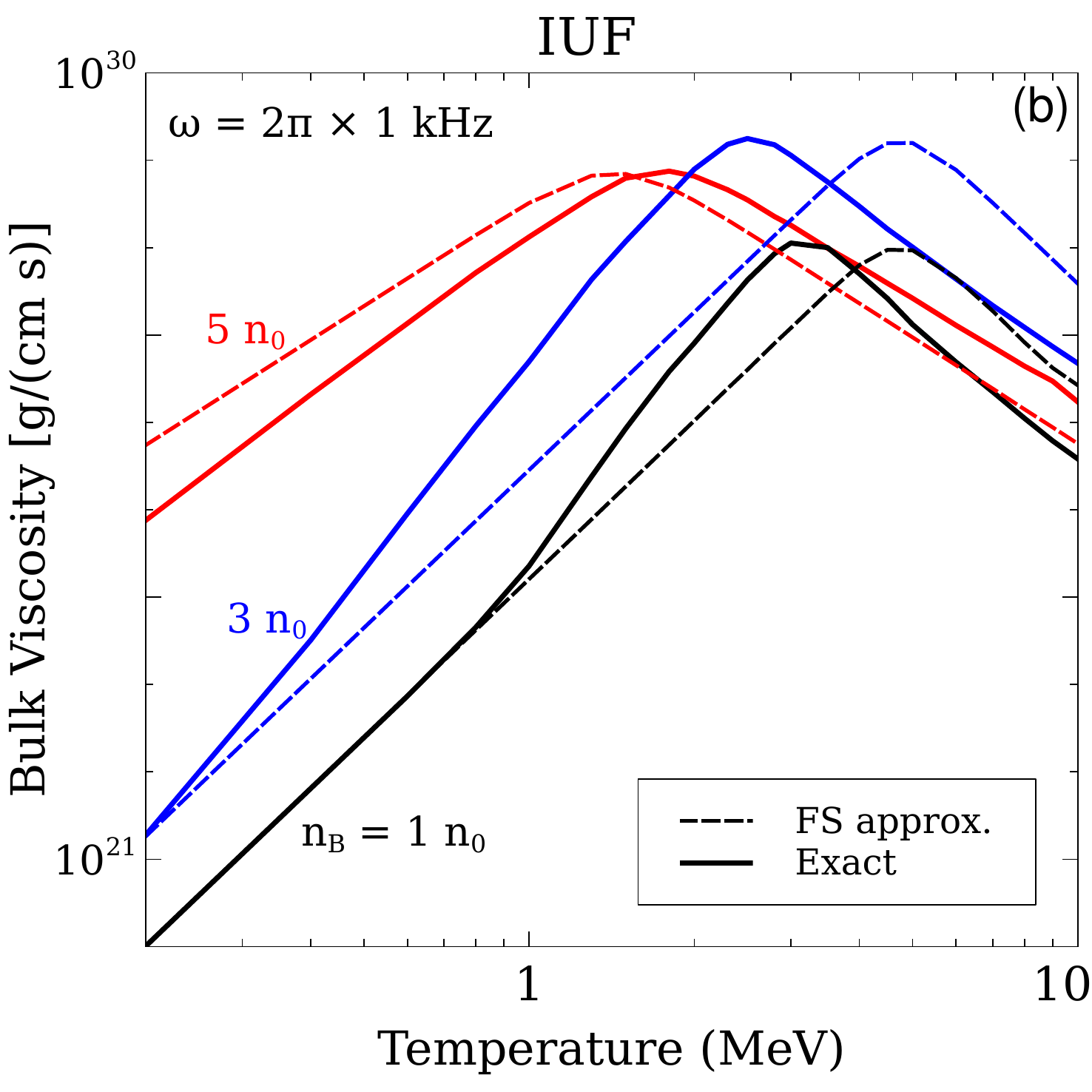}
    \end{subfigure}
    \caption{Bulk viscosity of nuclear matter as a function of temperature, for densities of $n_0$, $3n_0$, $5n_0$ when undergoing a density oscillation at 1 kHz. The equation of state is DD2 (a) or IUF (b).  Thin, dotted lines are the Fermi Surface approximation.  Thick, solid lines use the exact Urca rates.}
    \label{fig:bv_2d}
\end{figure*}

In Fig.~\ref{fig:bv_2d}, we show the bulk viscosity of nuclear matter with the DD2 and IUF EoS, when subjected to a 1 kHz density oscillation, which is a typical frequency for neutron star mergers \cite{Alford:2017rxf}.  The dashed lines are the bulk viscosity with Urca rates calculated in the Fermi Surface approximation while the solid lines use the exact Urca rates.  

In Fig.~\ref{fig:bv_2d}(a), corresponding to the DD2 EoS, the exact bulk viscosity peaks at a  temperature that is 1-2\,\MeV\ lower than would be predicted by the Fermi Surface approximation.  
This is because DD2 never allows direct Urca (the threshold is at infinite density), and we know (see Fig.~\ref{fig:urca_rates}) that the Fermi Surface approximation underestimates the below-threshold Urca rate.
This means that in the Fermi Surface approximation the temperature must be pushed up to a higher value in order for the equilibration rate to match the oscillation frequency, which is where the resonant peak occurs (Sec.~\ref{sec:bulk-visc-basics}).

In Fig.~\ref{fig:bv_2d}(b), corresponding to the IUF EoS, which has a direct Urca threshold near $4n_0$.
Here we see two distinct behaviors.  For densities $n_0$ and $3 n_0$, which are below threshold, the behavior is similar to that seen for DD2: the Fermi Surface approximation only includes modified Urca processes, but the exact calculation includes below-threshold direct Urca processes which increase the total rate, moving the resonant peak to lower temperatures.  Above the threshold density, the Fermi Surface approximation for direct Urca overestimates the total Urca rate, since the exact phase space integration leads to only a gradual opening of the phase space around the direct Urca threshold, hence the resonant peak moves to higher temperatures than predicted by the Fermi Surface approximation.

As can be seen from Eq.~\ref{eq:bv}, the maximum value of bulk viscosity at a frequency $\omega$ is 
\begin{equation}
    \zeta_{\text{max}} = \frac{C^2}{2B\omega}.
    \label{eq:bv-max}
\end{equation}
\begin{figure*}[t!]
    \centering
    \begin{subfigure}[t]{0.5\textwidth}
        \centering
        \includegraphics[width=.95\textwidth]{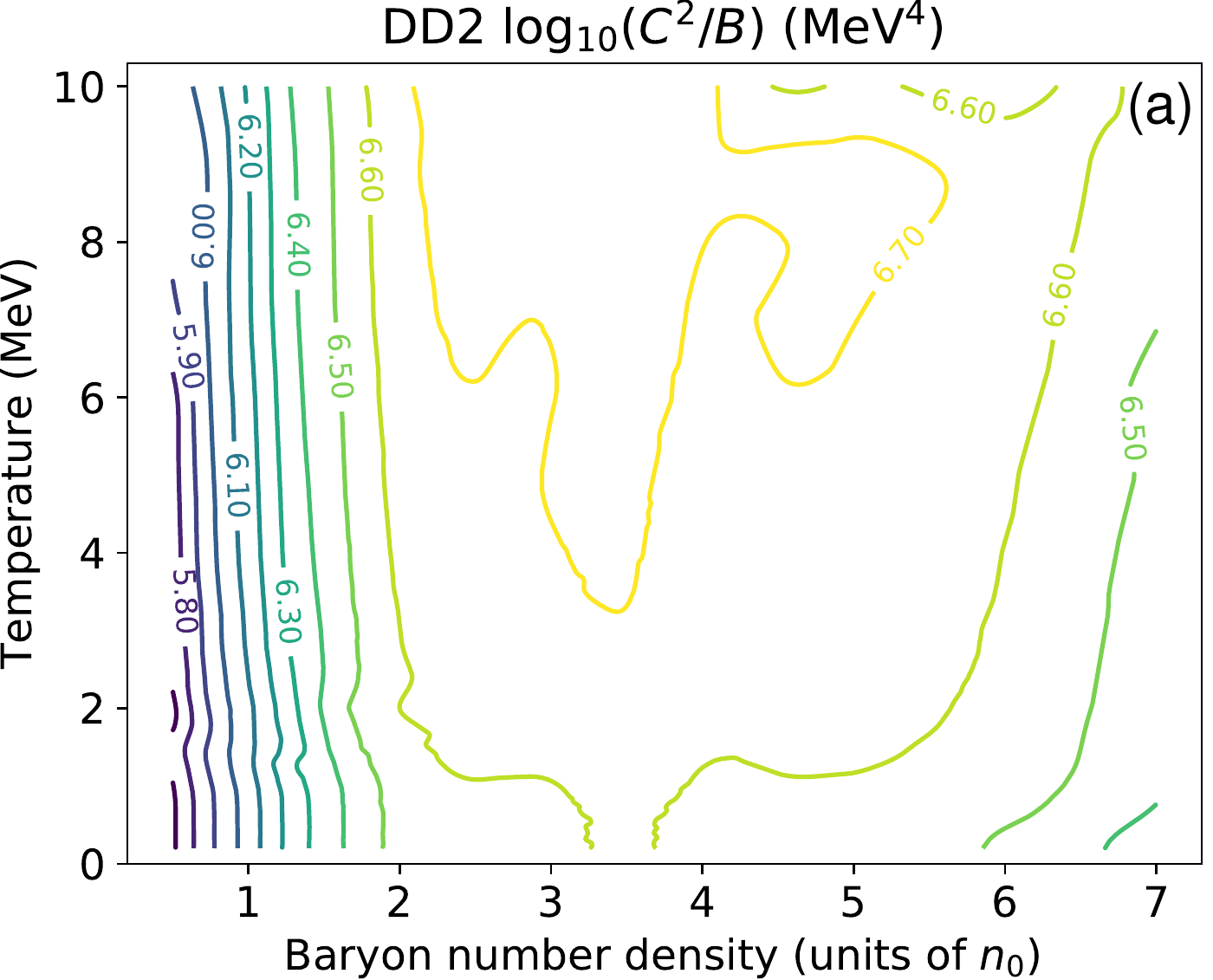}
    \end{subfigure}%
    \begin{subfigure}[t]{0.5\textwidth}
        \centering
        \includegraphics[width=.95\textwidth]{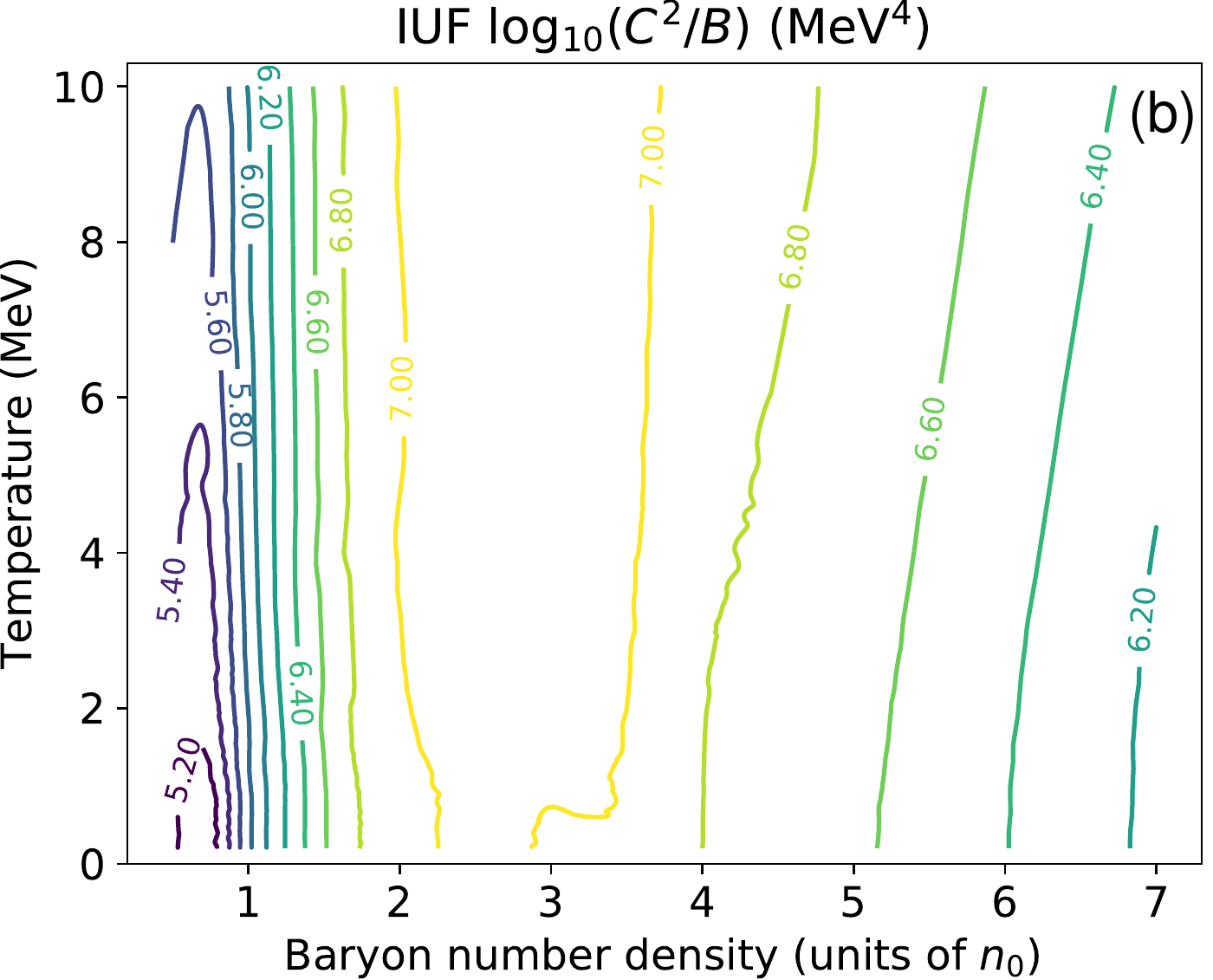}
    \end{subfigure}
    \caption{Logarithmic plot of the ratio of susceptibilities $C^2/B = 2\omega \zeta_{\text{max}}$ that determines the maximum bulk viscosity at a given oscillation frequency. We show results for the DD2 (a) and IUF (b) equations of state, calculated in beta equilibrium (Eq.~\ref{eq:full_beq}). }
    \label{fig:c2b}
\end{figure*}

In Fig.~\ref{fig:c2b} we plot $C^2/B = 2\omega\zeta_{\text{max}}$ for a representative range of densities and temperatures for which nuclear matter is likely neutrino-transparent.  We see that for a given frequency, the maximum value of bulk viscosity varies by 1-2 orders of magnitude, and depends more strongly on density than on temperature.  Most notably, we can see that $C^2/B$ rises rapidly at low densities, then levels off at $n\sim 2 n_{\rm sat}$ to a value about an order of magnitude larger then its value at $n=n_{\rm sat}$. (This could already be seen in Fig.~\ref{fig:bv_2d}).

\begin{figure*}[t!]
    \centering
    \begin{subfigure}[t]{0.5\textwidth}
        \centering
        \includegraphics[width=.95\textwidth]{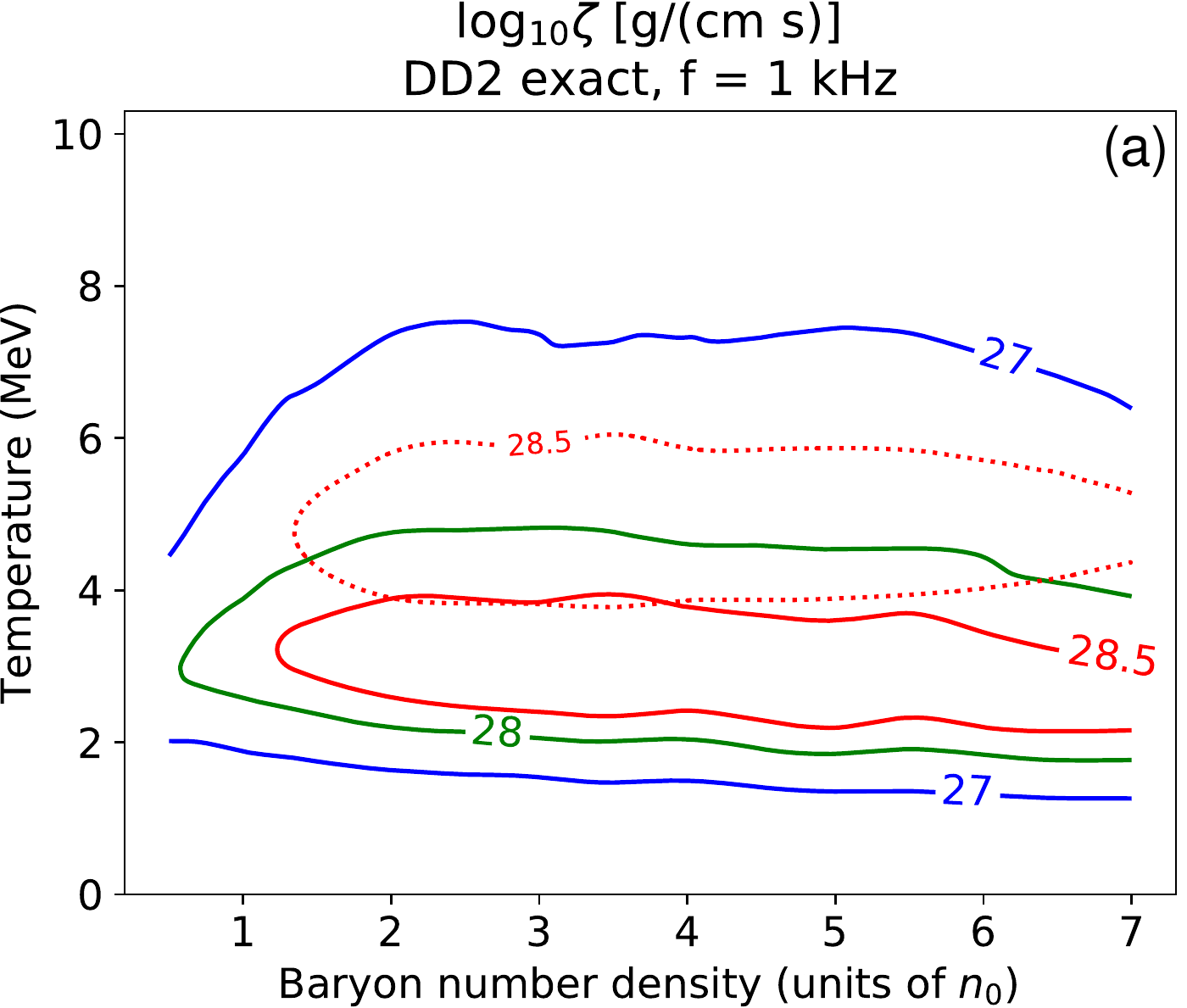}
    \end{subfigure}%
    \begin{subfigure}[t]{0.5\textwidth}
        \centering
        \includegraphics[width=.95\textwidth]{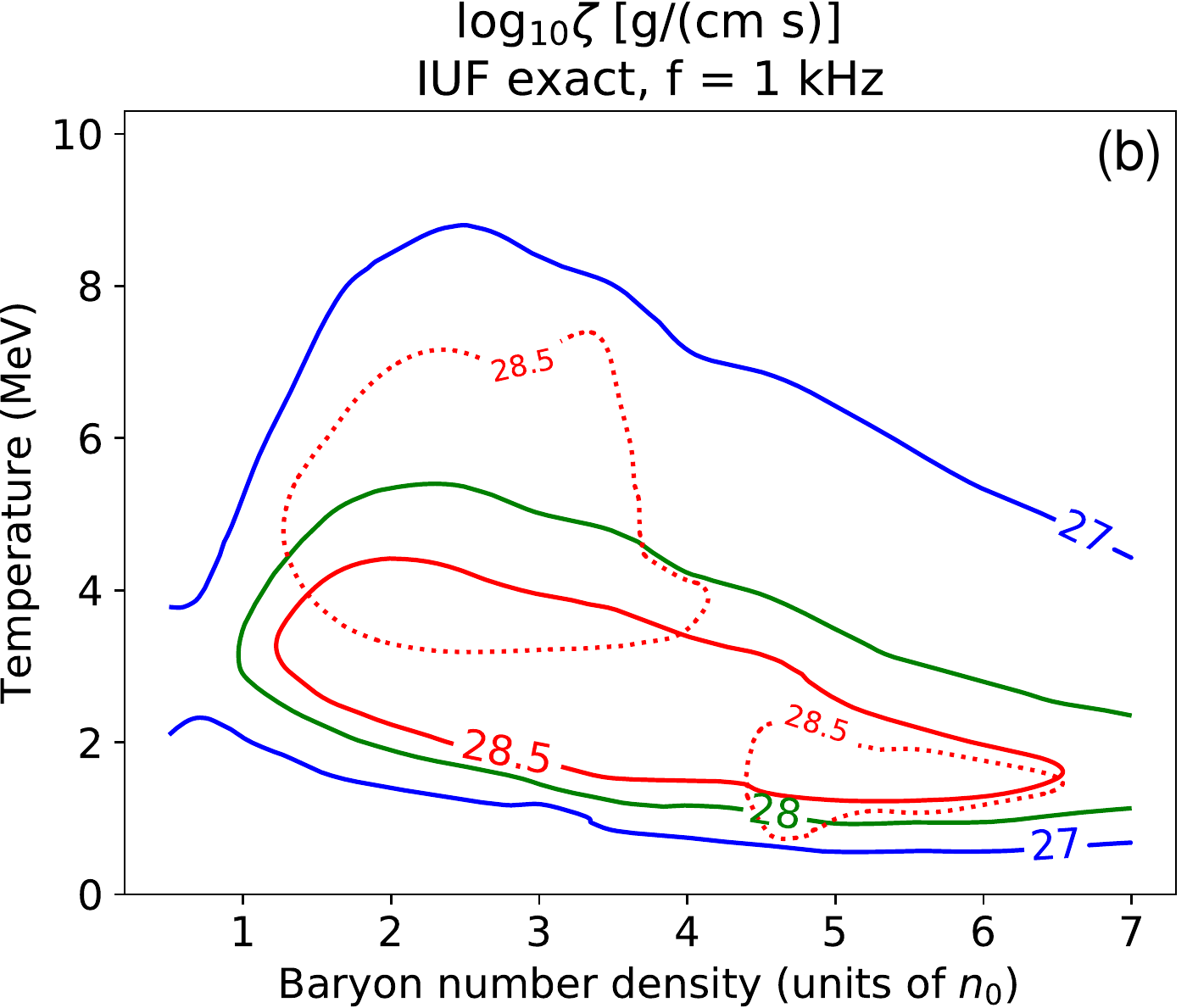}
    \end{subfigure}
    \caption{Bulk viscosity as a function of density and temperature, for the DD2 (a) and IUF (b) EoSs.  The full phase space integral for the direct Urca rate is used in the solid line contours, while the 28.5 dashed contour uses the FS approximation.}
    \label{fig:bv3d}
\end{figure*}

In Fig.~\ref{fig:bv3d}, we plot the bulk viscosity as a function of density and temperature (the curves in Fig.~\ref{fig:bv_2d} are cross-sections through Fig.~\ref{fig:bv3d}).  For a fixed density, as the temperature rises, the beta equilibration rate $\gamma$ rises rapidly because of the increase in available phase space.  At temperatures of a few MeV, the reequilibration rate closely matches the oscillation frequency of 1 kHz, then bulk viscosity reaches a maximum.  At higher temperatures, the reequilbration is too fast and the bulk viscosity drops.

We see that for the DD2 EoS, the bulk viscosity peak is at a temperature of about 3 MeV for all densities, which is a lower temperature than predicted by the Fermi surface approximation.  For IUF, the FS approximation would suggest two different peaks in bulk viscosity: one below the direct Urca threshold corresponding to the near-equality of the modified Urca rate and the density oscillation frequency, and one above the threshold, corresponding to the near-equality of the direct Urca rate and the density oscillation frequency.  However, the gradual opening of the direct Urca threshold coming from the exact direct Urca calculation melds these two peaks into one broad peak.  At low density, the peak is at 3-4 MeV, but as density increases it moves down to 2 MeV.

\subsection{Energy dissipation time}
The most direct indicator of the importance of bulk viscous damping is the dissipation time $\tau_{\rm diss}$ for density oscillations. Since the merging stars settle down into a massive remnant in tens of milliseconds, 
bulk viscous damping will be important if $\tau_{\rm diss}$ is tens of milliseconds or less.
To calculate the dissipation time, we need the energy of an oscillation and the rate at which that energy  is 
dissipated by bulk viscosity.
The energy density of an adiabatic baryon density oscillation $n_B(t)=n_B + (\Delta n)\sin{(\omega t)}$ is \cite{Alford:2017rxf}
\begin{equation}
    \varepsilon = \frac{1}{2} (\Delta n)^2  \frac{\partial^2\varepsilon}{\partial n_B^2}\bigg\vert_{x_p,s/n_{B}} 
    = \frac{\kappa_S^{-1}}{2}\left(\frac{\Delta n}{n_B}\right)^2 \ ,
    \label{eq:osc-energy}
\end{equation}
where $\kappa_S$ is the adiabatic compressibility\cite{schroeder1999introduction,compose_manual}
\begin{equation}
    \kappa_S^{-1} = n_B \frac{\partial P}{\partial n_B}\bigg\vert_{x_p,s/n_B} \ .
\end{equation}
See Sec.~\ref{sec:bulk-visc-basics} for a discussion
of the assumption of adiabaticity.

\begin{figure}
\includegraphics[width=.45\textwidth]{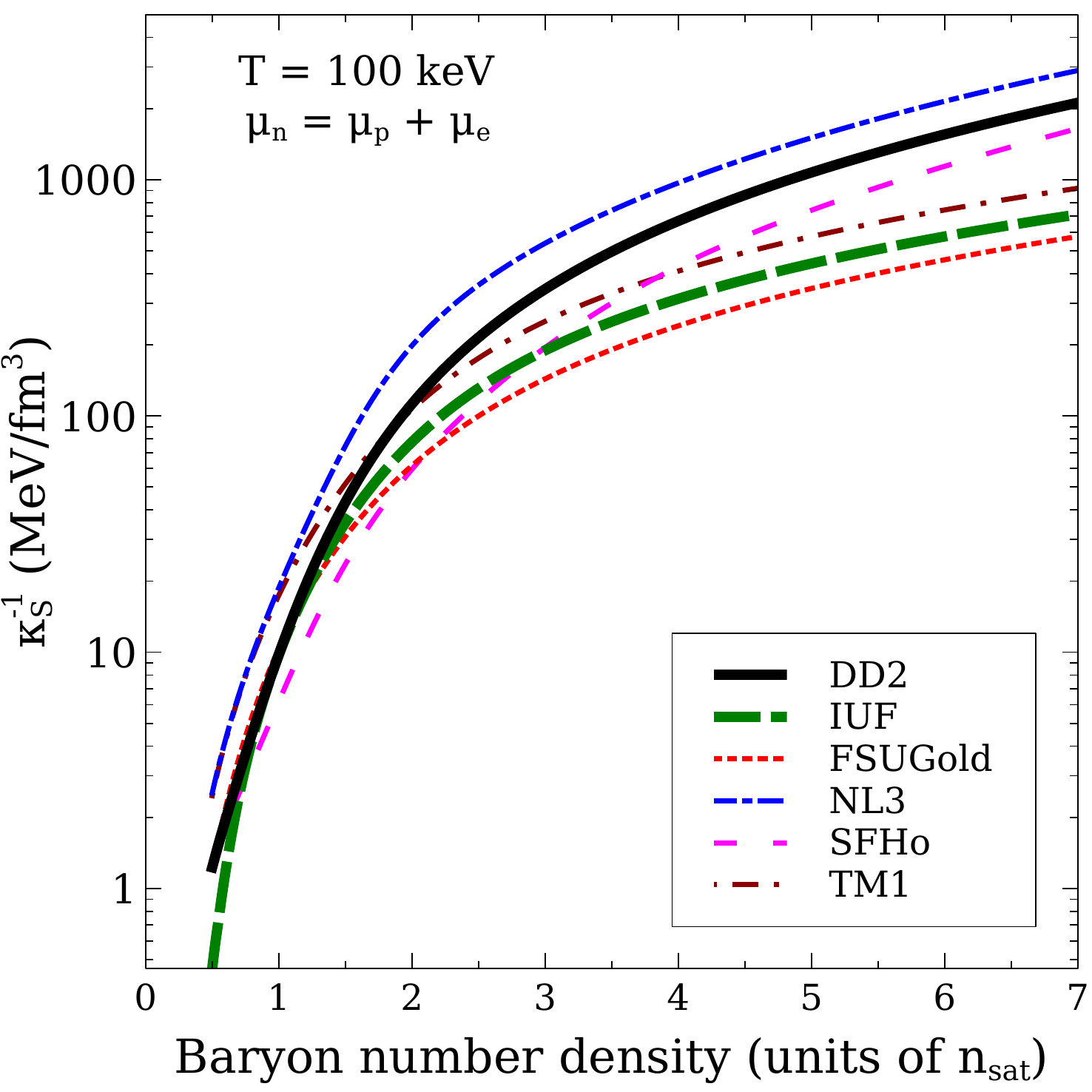}
\caption{ Adiabatic inverse compressibility $(\kappa_S)^{-1}$ at low temperature, versus density, for several EoSs derived from relativistic mean-field theories. 
Each is tabulated on CompOSE.}
\label{fig:K}
\end{figure}

We note that the adiabatic compressibility depends on the EoS.
However, it is a common feature of all nucleonic EoSs that  nuclear matter becomes more incompressible at high densities,
so the inverse compressibility  $1/\kappa_S$ rises with density, as shown in Fig.~\ref{fig:K} for a range of EoSs including those used in this work. This means that at higher density, oscillations in the density store more energy.  

To facilitate comparison with previous work (for example, \cite{Alford:2017rxf}), we mention that the ``stiffness'' of nuclear matter is often described via the nuclear incompressibility $K$ \cite{Shapiro:1983du,glendenning2000compact,Schmitt:2010pn}.  $K$ is conventionally defined at saturation density, zero temperature, and for symmetric nuclear matter, and is approximately 250\,MeV~\cite{Oertel:2016bki}.  Some works have extended the definition of the nuclear incompressiblity to densities above nuclear saturation \cite{Dexheimer:2007mt}.  At zero temperature, $\nsat$, and for symmetric nuclear matter, the adiabatic $\kappa_S$ can be related to the nuclear incompressibility $K$ by 
$K = 9/(\kappa_S n_0)$~\cite{compose_manual,Schmitt:2010pn}.

The rate of energy density dissipation is given by \cite{Alford:2010gw,Sawyer:1980wp} 
\begin{equation}
\frac{\mathop{d\varepsilon}}{\mathop{dt}} = \frac{\omega^2}{2}\left(\!\frac{\Delta n}{n_B}\!\right)^{\!2} \zeta \ . 
\label{eq:osc-dissipation}
\end{equation}
Using Eq.~\ref{eq:osc-energy}, the energy dissipation time is 
\begin{equation}
    \tau_{\text{diss}} \equiv \frac{\varepsilon}{d\varepsilon/dt} = \frac{(\kappa_S)^{-1}}{\omega^2 \zeta} \ .
    \label{eq:t-diss}
\end{equation}
Note that one can also define \cite{1990ApJ...363..603C} a decay time for the amplitude, which would
be longer by a factor of two since the energy of an oscillation goes as the square of the amplitude.  
\begin{figure*}[t!]
    \centering
    \begin{subfigure}[t]{0.5\textwidth}
        \centering
        \includegraphics[width=.95\textwidth]{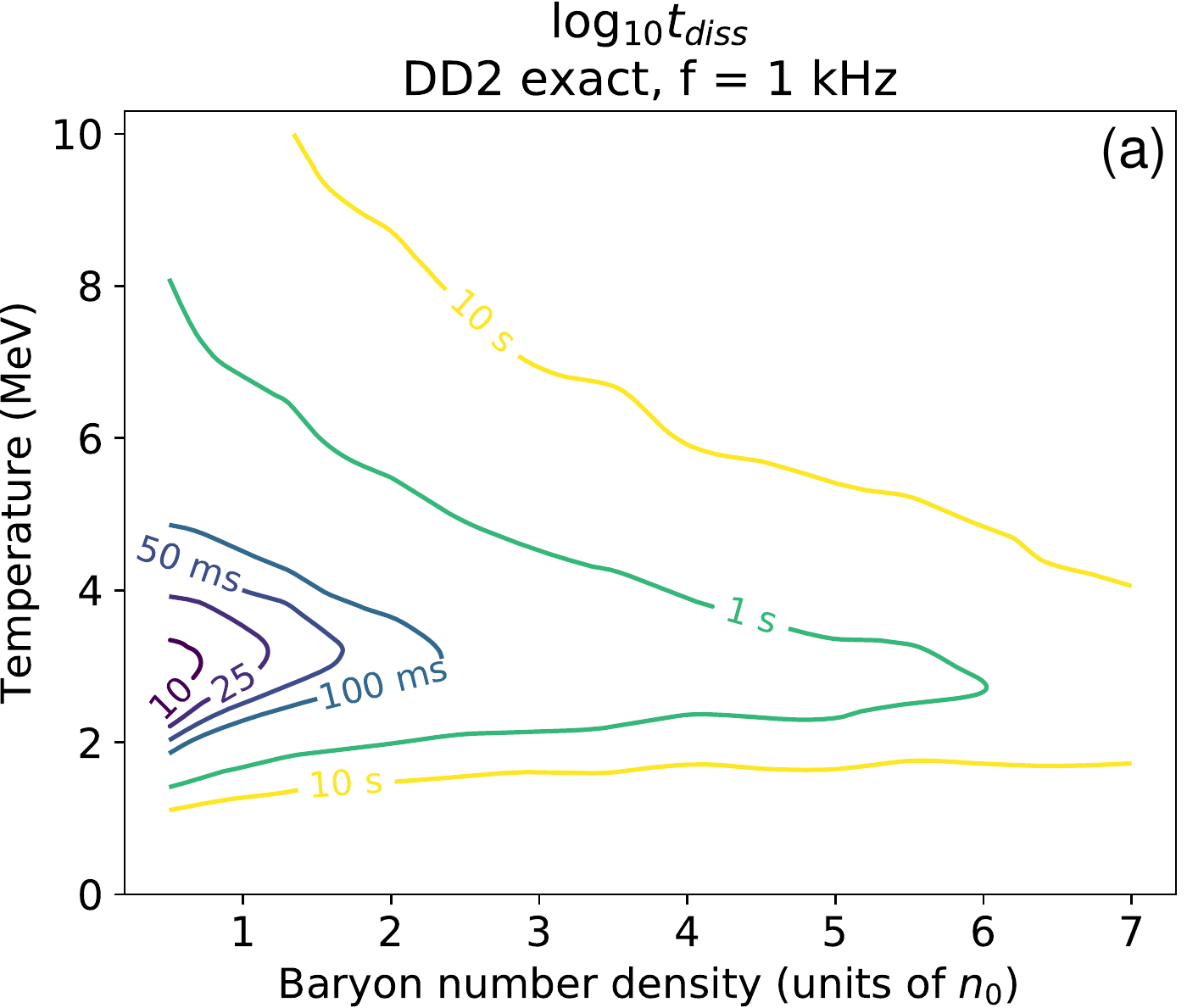}
    \end{subfigure}%
    \begin{subfigure}[t]{0.5\textwidth}
        \centering
        \includegraphics[width=.95\textwidth]{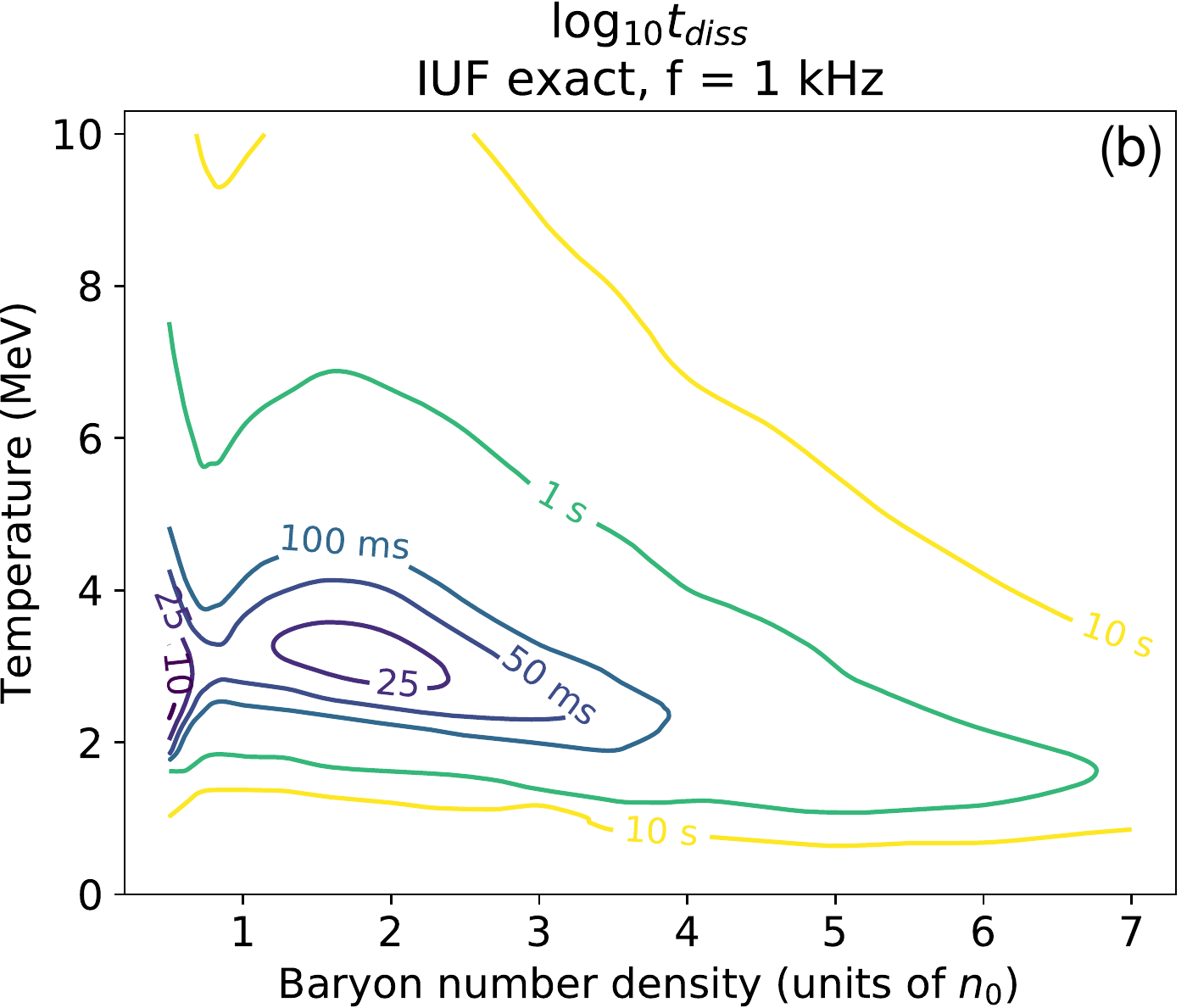}
    \end{subfigure}
    \caption{Dissipation time $\tau_{\rm diss}$ of a 1\,kHz density oscillation, using the DD2 EoS (a) and IUF EoS (b), with the exact Urca rates.}
    \label{fig:tdiss_1kHz_exact}
\end{figure*}
\begin{figure*}[t!]
    \centering
    \begin{subfigure}[t]{0.5\textwidth}
        \centering
        \includegraphics[width=.95\textwidth]{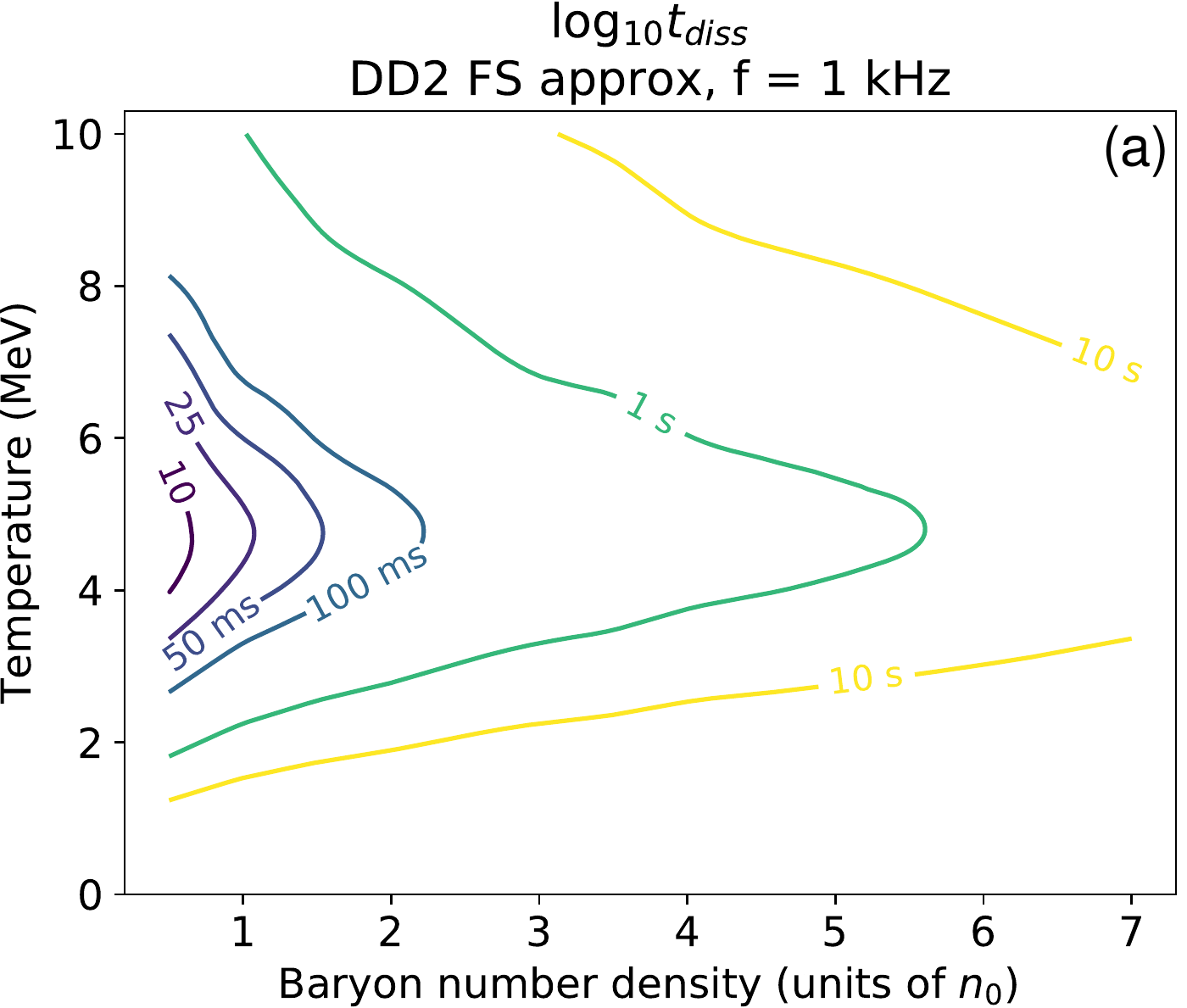}
    \end{subfigure}%
    \begin{subfigure}[t]{0.5\textwidth}
        \centering
        \includegraphics[width=.95\textwidth]{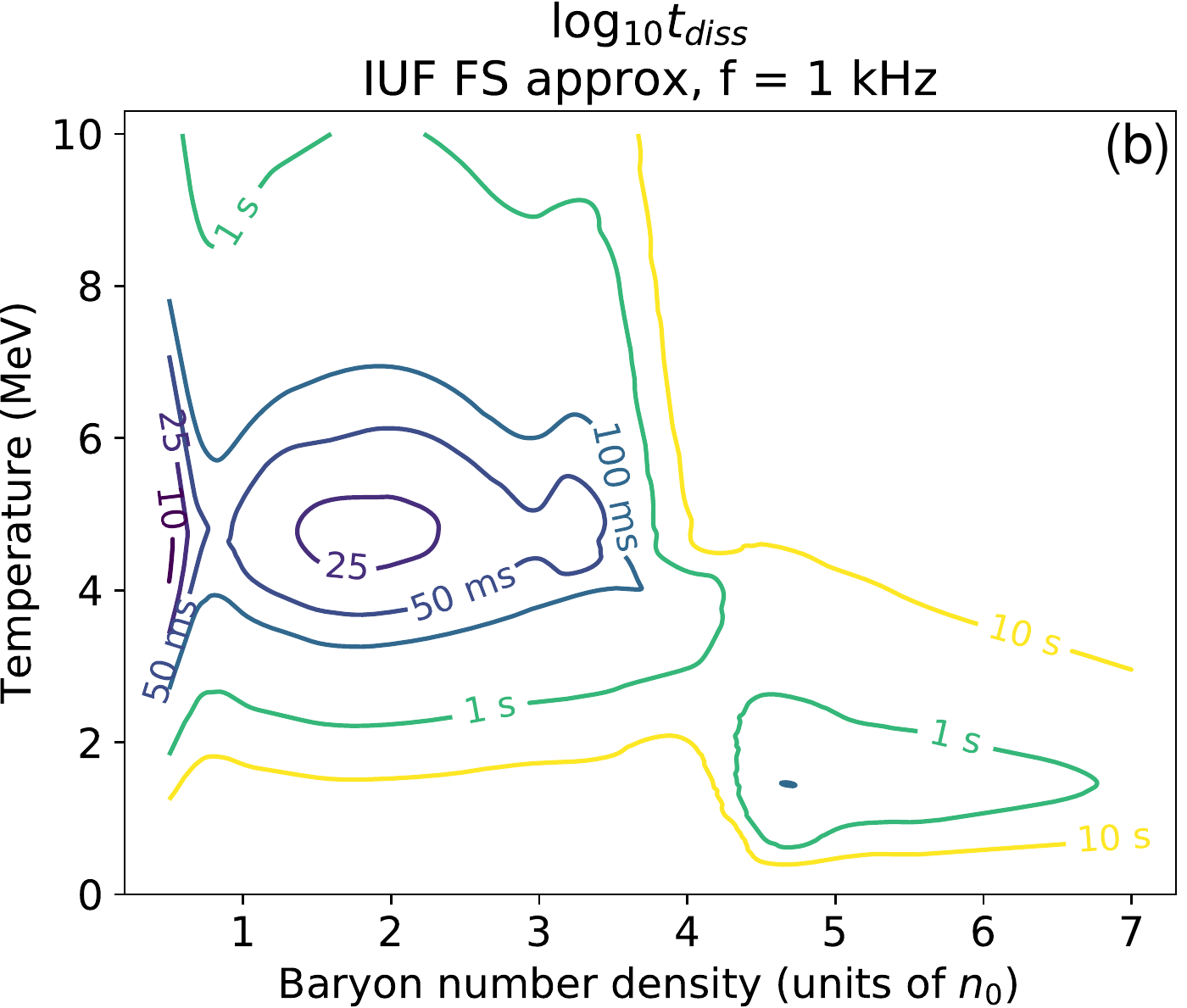}
    \end{subfigure}
    \caption{Dissipation time $\tau_{\rm diss}$ of a 1\,kHz density oscillation, using the DD2 EoS (a) and the IUF EoS (b), calculated in the Fermi Surface approximation.}
    \label{fig:tdiss_1kHz_FS}
\end{figure*}

In Fig.~\ref{fig:tdiss_1kHz_exact}, we plot the dissipation time of a 1\,kHz density oscillation as a function of density and temperature for two different EoSs, using the exact Urca rates.  We first discuss the physical content and implications of the exact results (Fig.~\ref{fig:tdiss_1kHz_exact}), then compare them to the Fermi Surface approximation, shown in Fig.~\ref{fig:tdiss_1kHz_FS}.

\noindent {\em Temperature dependence}. The adiabatic compressibility is relatively independent of temperature, so the bulk viscosity dominates the temperature dependence of the dissipation time.  As discussed in Sec.~\ref{sec:bv_results}, for a given density, the bulk viscosity increases, reaches a resonant maximum when the beta reequilibration rate $\gamma$ matches the oscillation frequency $\omega$, and then decreases as temperature increases.  This leads to minimum dissipation time at approximately the temperature at which the bulk viscosity reaches its maximum, for a given density.

\noindent {\em Density dependence}. 
The adiabatic inverse compressibility strongly increases as a function of density, as seen in Fig.~\ref{fig:K}.  While the bulk viscosity was weakly dependent on density, the dissipation time at high density is strongly increased due to the several order-of-magnitude rise of the adiabatic inverse compressibility.  Physically, oscillations in high density nuclear matter have a lot of energy due to the high incompressiblity of dense nuclear matter (see Eq.~\ref{eq:osc-energy}).  Thus, it takes correspondingly longer time for those high energy oscillations to damp.  As a result of the behavior of the compressibility of nuclear matter, the minimum of dissipation time is likely to be located at a low density.

It is worth noting that the bulk viscosity varies non-monotonically with density.  It rises as density increases from $0.5\nsat$, reaches a peak at several times $\nsat$, and then falls off at high density.  This can be seen by noting that the maximum bulk viscosity is $\zeta_{\rm max} = (1/2\omega) C^2/B$ (Eq.~\ref{eq:bv-max}), which is plotted in Fig.~\ref{fig:c2b}.  It is clear that the particular features of the rise and fall in bulk viscosity as a function of density depend on the EoS.  Throughout the range of densities that we consider, the bulk viscosity prefactor $C^2/B$ varies by 1-2 orders of magnitude.  However, the inverse compressibility rises by three orders of magnitude over that density range, so it has a more substantial effect on the density dependence of the dissipation time.

For the DD2 EoS, as seen in Fig.~\ref{fig:tdiss_1kHz_exact}(a), the minimum dissipation times lie around temperatures of 3 MeV for all densities, indicating that the reequilibration rate doesn't change strongly with density, which is expected since only modified Urca and below-threshold direct Urca are acting.  As a function of density, the dissipation times get longer as density increases.  This behavior comes from the dramatic monotonic rise of the inverse compressibility as a function of density.  The bulk viscosity prefactor $C^2/B$ rises by one order of magnitude from $0.5\nsat$ to 3 or 4 $\nsat$, and then slightly decreases at higher densities, but it doesn't vary rapidly enough to compete with the rise of the inverse compressibility, and thus the dissipation time rises monotonically with density.  DD2 has a minimum dissipation time of about 6 ms, which occurs only at low density ($0.5\nsat$) at temperatures of just under 3 MeV.  Only fluid elements with densities under twice saturation density would dissipate energy on timescales relevant for mergers.  

As seen in Fig.~\ref{fig:tdiss_1kHz_exact}(b), the behavior of the dissipation time scale for the IUF EoS is more complicated.  The lowest dissipation times do occur at temperatures of around 3 MeV, since the resonant peak of bulk viscosity is around that temperature.  However, the nonmonotonic behavior of $C^2/B$ as a function of density is more dramatic for the IUF EoS than for DD2, so it competes with the rapidly rising inverse compressibility as density increases, leading to two minima in the dissipation time.  The first is at low density, where the nuclear inverse compressibility is decreasing rapidly as the density decreases to the lowest value for which we trust our equation of state, $n=0.5\nsat$. There, energy dissipation can occur in as little as 5 ms.  There is also a local minimum around $n=2\nsat$, where the bulk viscosity prefactor $C^2/B$ has a local maximum (see Fig.~\ref{fig:c2b}(b)) and dissipation times reach down to 19 ms.   For the IUF EoS, dissipation occurs on merger timescales in fluid elements up to four times saturation density, in contrast to the behavior of DD2.

It is interesting to compare the Fermi Surface approximate results (Fig.~\ref{fig:tdiss_1kHz_FS}) and the exact results (Fig.~\ref{fig:tdiss_1kHz_exact}) for each EoS.  For DD2, the use of the exact Urca rates just increases the total Urca rate and thus the bulk viscosity is maximized at a lower temperature than would be predicted by the Fermi Surface approximation.  For IUF, the Fermi Surface approximate result would predict a sharp change in the behavior of the bulk viscosity at the direct Urca threshold,  $n=4n_{\text{sat}}$ (for a generic example of this behavior, see Figure 1 in \cite{Haensel:2001mw}).  However, at the temperatures of interest to us the exact Urca rates show a gradual increase with density and thus the bulk viscosity does not change suddenly at the threshold density.    
\subsection{Higher frequency oscillations}
\begin{figure*}[t!]
    \centering
    \begin{subfigure}[t]{0.5\textwidth}
        \centering
        \includegraphics[width=.95\textwidth]{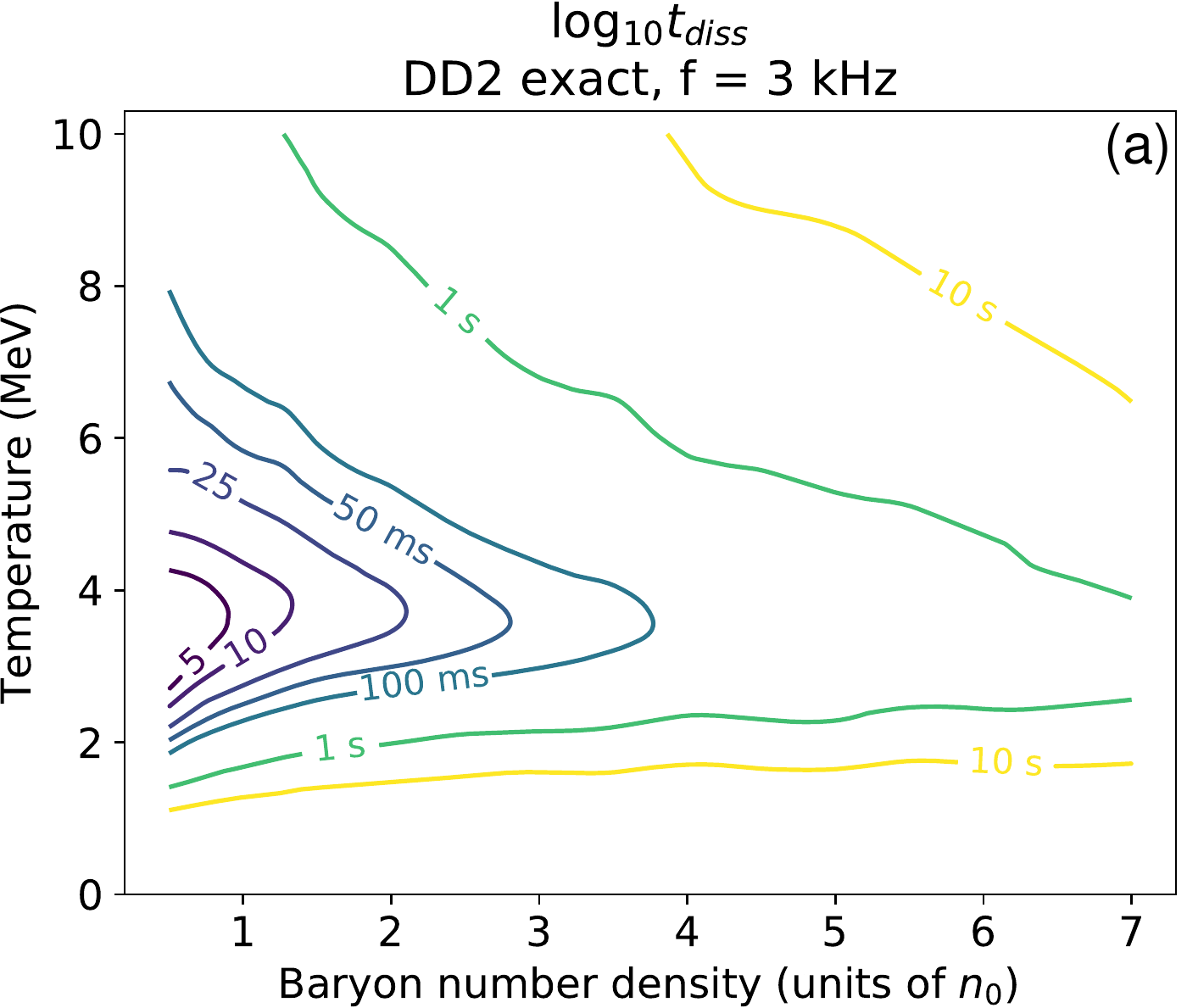}
    \end{subfigure}%
    \begin{subfigure}[t]{0.5\textwidth}
        \centering
        \includegraphics[width=.95\textwidth]{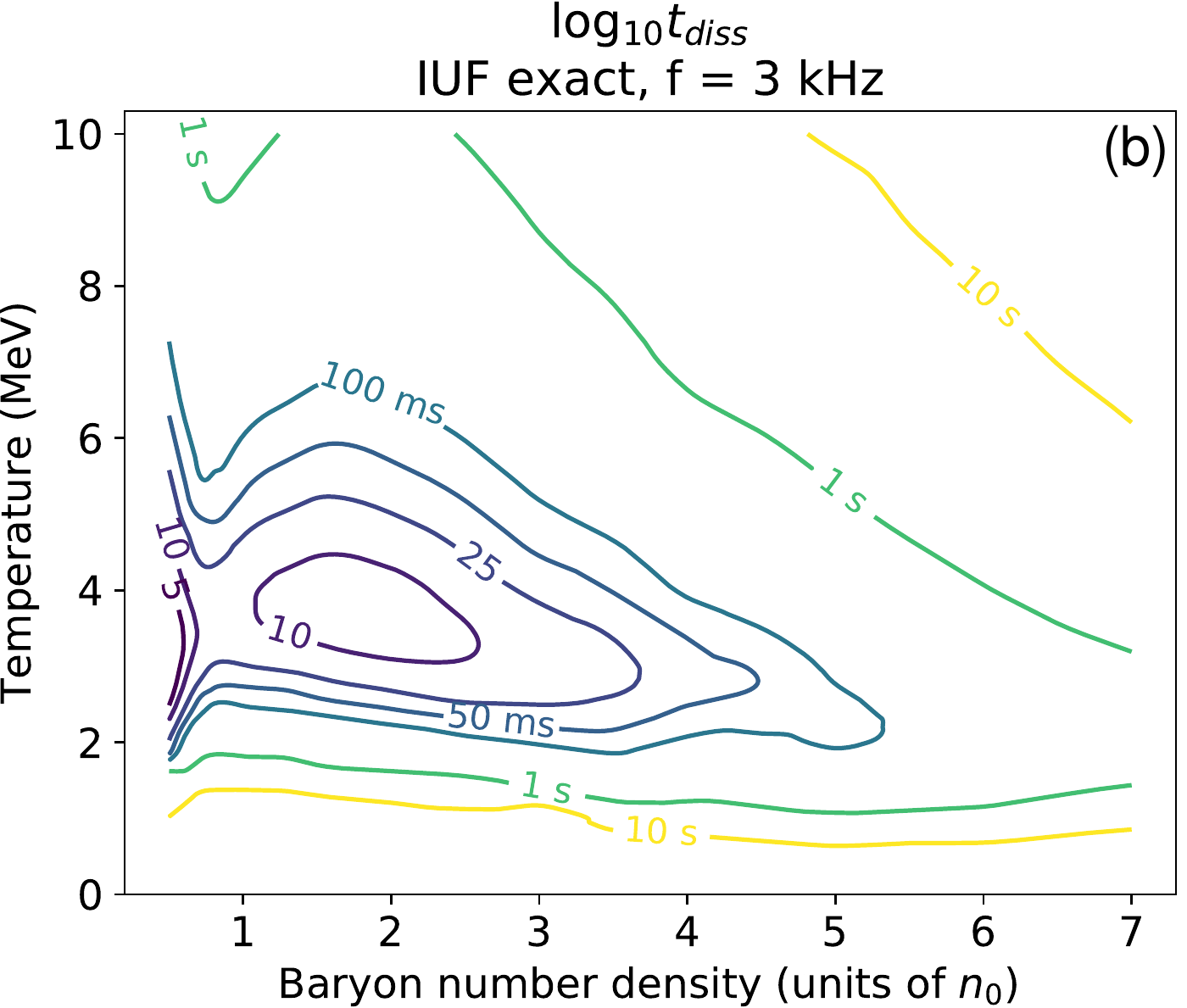}
    \end{subfigure}
    \caption{Dissipation time scale for 3 kHz oscillations in nuclear matter with the DD2 EoS (a) or IUF EoS (b).  The exact Urca rates are used.}
    \label{fig:3kHz}
\end{figure*}
\begin{figure*}[t!]
    \centering
    \begin{subfigure}[t]{0.5\textwidth}
        \centering
        \includegraphics[width=.95\textwidth]{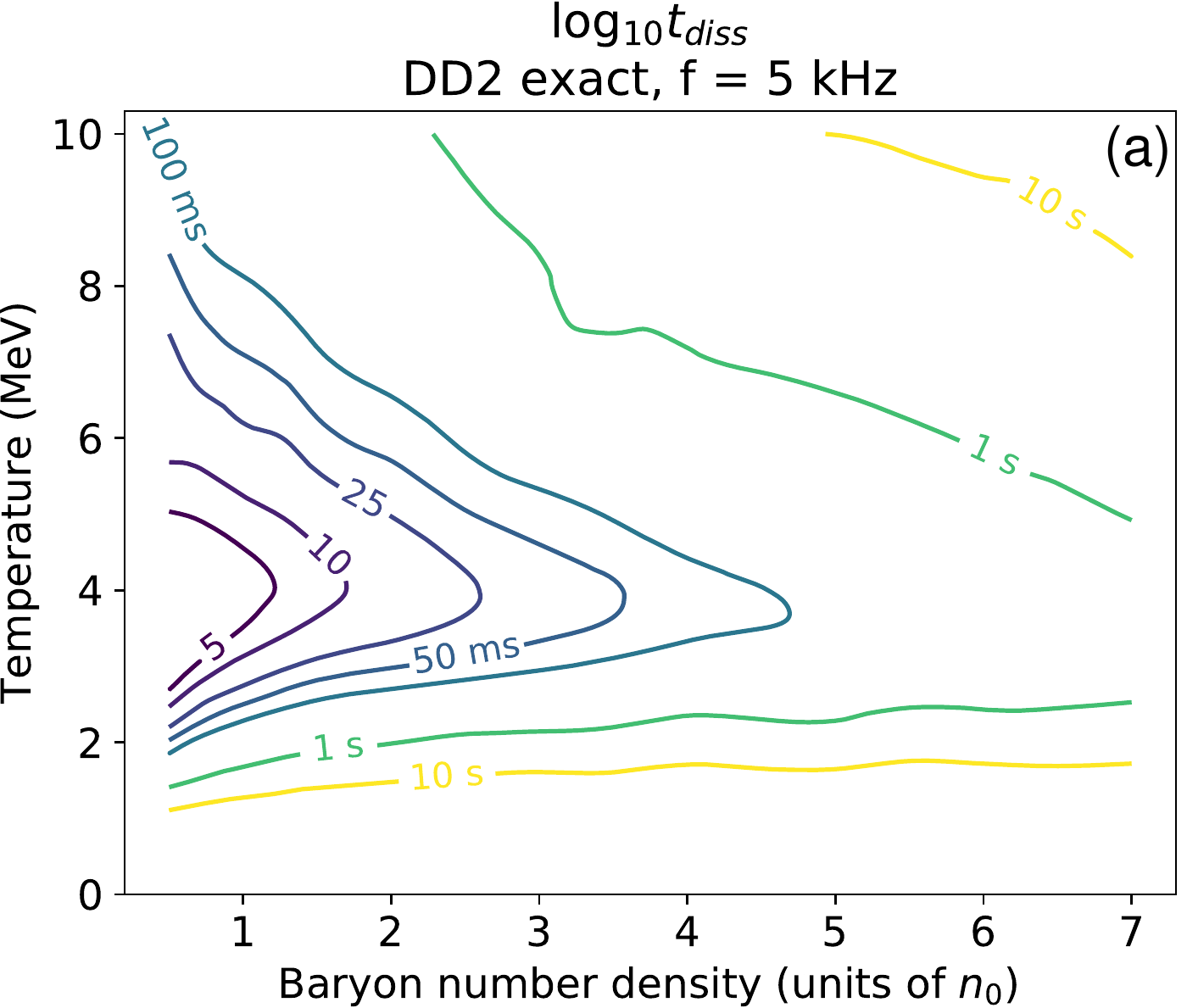}
    \end{subfigure}%
    \begin{subfigure}[t]{0.5\textwidth}
        \centering
        \includegraphics[width=.95\textwidth]{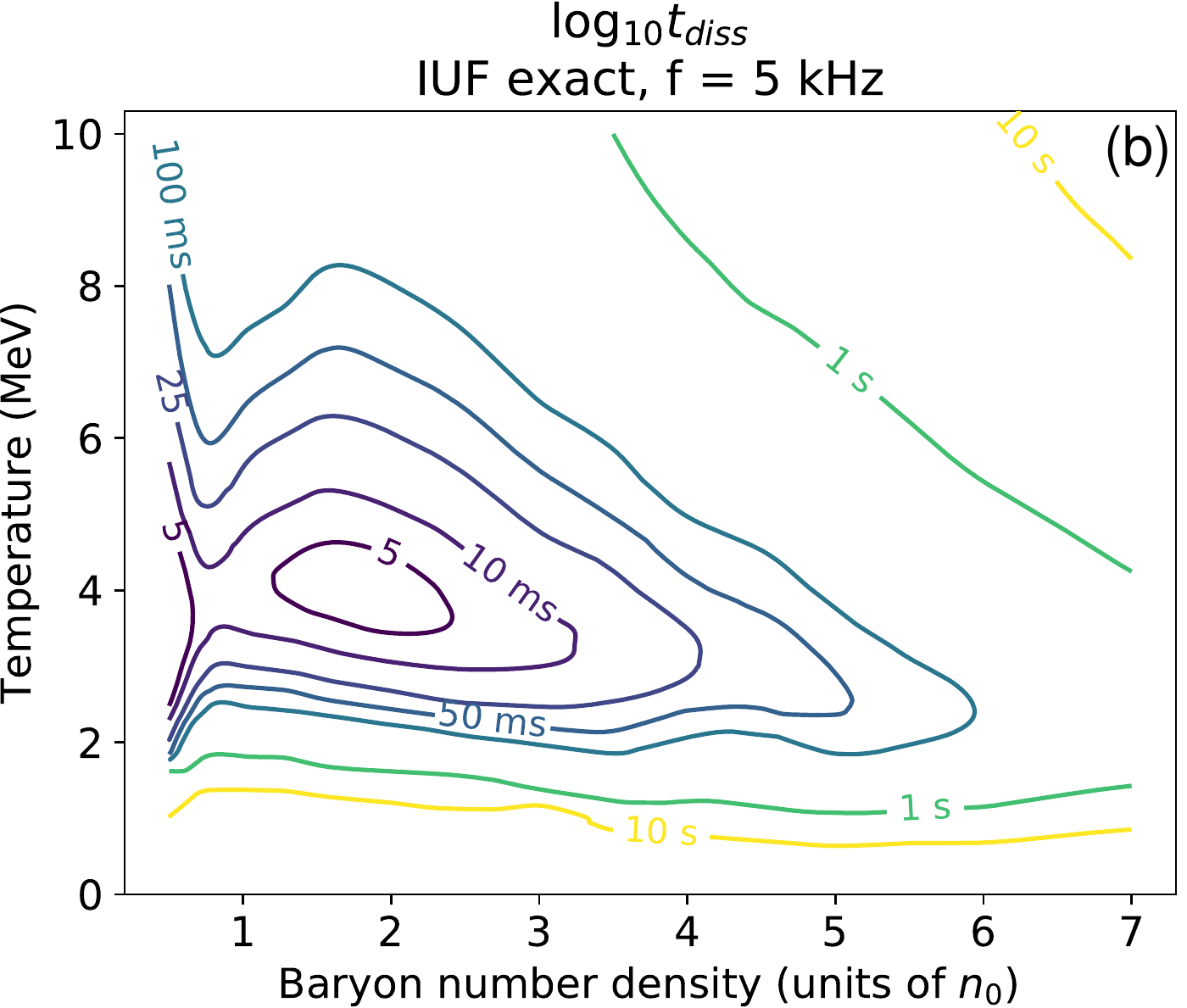}
    \end{subfigure}
    \caption{Dissipation time scale for 5 kHz oscillations in nuclear matter with the DD2 EoS (a) or IUF EoS (b).  The exact Urca rates are used.}
    \label{fig:5kHz}
\end{figure*}
There is evidence from simulations \cite{Hotokezaka:2013iia,Chaurasia:2018zhg,Clark:2015zxa,Rezzolla:2016nxn,Dietrich:2016hky,Maione:2017aux,Bauswein:2018bma} (see also the review \cite{Baiotti:2019sew}) that eccentric binary neutron star mergers excite oscillations at frequencies above 1 kHz.  We plot the dissipation times for 3 kHz and 5 kHz oscillations in Figs \ref{fig:3kHz} and \ref{fig:5kHz}.  We see that at these higher frequencies, bulk viscosity plays a bigger role, and density oscillations can be damped in as little as 1 ms, and for a broad range of temperatures and densities, oscillations can be damped in under 25 ms.

We note that, at a given density, a higher temperature is required to make the reequilibration rate $\gamma$ match a density oscillation which has a frequency above 1 kHz, and thus the region of maximum bulk viscosity is moved to higher temperatures.  For example, a 5 kHz density oscillation has maximum bulk viscosity (and thus minimum damping time) at about $T = 4 \text{ MeV}$ (see Fig.~\ref{fig:5kHz}), while a 1 kHz density oscillation has maximum bulk viscosity at around $T = 3 \text{ MeV}$ (see Fig.~\ref{fig:tdiss_1kHz_exact}).
\section{Conclusions}

We have calculated the bulk-viscous dissipation time in nuclear matter at temperatures and densities relevant to neutron star mergers. 
We assumed the material was transparent to neutrinos, which should be valid for temperatures up to about 5\,MeV, and we studied the damping of oscillations with frequencies in the 1\,kHz range, which are seen in simulations of mergers.
The main uncertainty in our result is the form of the
nuclear matter equation of state at supranuclear densities,
so we performed calculations for two different equations of state, one stiffer, DD2, and one softer, IUF, which differ in their treatment of the nucleon-meson interaction (see
Sec.~\ref{sec:models}). 
Our main results are displayed in
Fig \ref{fig:tdiss_1kHz_exact}.

Bulk viscous damping will play a significant role at densities and temperatures where the dissipation time is comparable to or less than the typical timescale of the merger, which is in the range of tens of milliseconds. Both equations of state show a similar overall pattern: bulk viscosity damps oscillations on timescales comparable to a merger for nuclear matter at temperatures of 2-4\,MeV and for densities between $0.5\nsat$ to $2\nsat$, with IUF also exhibiting fast damping for densities up to $4\nsat$.  Both EoSs have minimum dissipation times of about 5 ms, occurring at $0.5\nsat$, while IUF has another local minimum of dissipation time, about 20 ms, occurring at $2\nsat$.  The occurrence of dissipation times in the 10\,ms range leads us to conclude that bulk viscous damping should be seriously considered for inclusion in future simulations.
The strong dissipation that we see at low density may be relevant to the density oscillations that were found to be associated with mass ejection in the outer regions of the merger (See Fig.~9 of Ref.~\cite{DePietri:2019khb}).

There are several directions in which this topic could be further developed. In parallel with this work, an analogous calculation of the bulk viscosity for nuclear matter with trapped neutrinos was conducted \cite{Alford:2019kdw}.  This is appropriate for nuclear matter 
at temperatures well above 5\,MeV, and is relevant to mergers
because temperatures up to 80 or even 100\,MeV are predicted by simulations. 
It should be noted that the neutrino-transparent and neutrino-trapped regimes are the simplified extremes of a continuum, the whole of which is probably realized at different regions and stages of a merger.  Between these
extremes lies the regime where the spectrum of neutrinos
includes a low energy population that escapes, a high energy
tail that is trapped, and an intermediate energy range where the mean free path is comparable to the distance scale of the
fluid flows, requiring explicit inclusion of neutrinos in
the dynamics of the nuclear fluid \cite{Ardevol-Pulpillo:2018btx,Perego:2014qda,Galeazzi:2013mia,Sekiguchi:2012uc,Rosswog:2003rv,1999JCoAM.109..281M}.

Another limitation of our calculation is the assumption of low-amplitude density oscillations. We calculated the ``subthermal'' bulk viscosity, but simulations show high amplitude density oscillations \cite{Alford:2017rxf} for which the suprathermal bulk viscosity \cite{Alford:2010gw} is relevant. This could extend the region of large bulk viscosity down to lower temperatures, since suprathermal effects allow high-amplitude oscillations to experience the maximum bulk viscosity $\zeta_{\rm max}$ (Eq.~\ref{eq:bv-max}) at lower temperatures \cite{Alford:2010gw}.

Our discovery of short bulk-viscous dissipation times at densities below nuclear saturation density, primarily due to the low inverse compressibility, underscores the need for a detailed understanding of the structure of nuclear matter below saturation density.  The DD2 and IUF EoSs predict uniform nuclear matter down to densities of 0.25 to 0.4 $\nsat$ respectively, which is why we restricted our calculations to densities  above $0.5\nsat$.  However,  a sequence of mixed ``pasta'' phases has been predicted at densities between $0.2$ and $0.7$ $\nsat$ \cite{Grill:2014aea,Fattoyev:2017zhb,Pais:2015lma,Pais:2014hoa,Oyamatsu:1993zz}.  It has been noted \cite{Yakovlev:2018jia, Gusakov:2004mj} that the appearance of free protons in certain pasta phases would open up the direct Urca process, albeit with such a reduced rate that it would take temperatures of tens of MeV---which is well above the pasta melting temperature of a few MeV \cite{Roggero:2017pag}---to reach the resonant peak of bulk viscosity.  Thus, it is important to know how and at what densities and temperatures nuclear matter transforms from a uniform phase to a mixed phase.  Based on our findings above, we expect subthermal bulk viscosity to be large for these low densities, down to the density at which uniform nuclear matter transitions to a pasta or spherical nuclei phase. 

As mentioned in section \ref{sec:nuclear}, we did not consider Urca processes involving muons, and did not include muons in the EoSs.  The presence of muon Urca processes would increase the equilibration rate $\gamma$ for densities at which muons are present.  In addition, muon-electron conversion would give rise to a separate contribution to the bulk viscosity \cite{Alford:2010jf}.  The calculation of bulk viscous damping time in Ref.~\cite{Alford:2017rxf} uses EoSs that contain muons. Above the onset density for muons the nuclear matter susceptibilities are larger, which would lead to larger bulk viscosity and thus shorter dissipation times compared to the muonless EoSs considered in this work. We are therefore planning to perform a full study of bulk viscous dissipation in EoSs that include muons. 

There is evidence that properly including in-medium effects in the nucleon propagator can lead to a large increase in the modified Urca rate just below the direct Urca threshold \cite{Shternin:2018dcn}.  We have not included this in our analysis, but it could potentially lead to a shift of the resonant peak of bulk viscosity to lower temperatures for a range of densities near the direct Urca threshold.

\section{Acknowledgments}
We thank Armen Sedrakian, Kai Schwenzer, Alex Haber, Reed Essick, Andreas Reisenegger, Alessandro Drago, and J\"{u}rgen Schaffner-Bielich for discussions.  We also thank the Frankfurt Institute for Advanced Study where part of this work was performed.  This research was partly supported by the U.S. Department of Energy, Office of Science, Office of Nuclear Physics, under Award No.~\#DE-FG02-05ER41375.

\appendix
\section{Adiabatic and isothermal oscillations}
\label{app:adiabatic}

Most previous works use the isothermal susceptibilities 
\begin{align}
 B_T &= -\frac{1}{n_B} \frac{\partial\mu_{\Delta}}{\partial x_p} \bigg\rvert_{n_B,T} \ , \\
 C_T &= n_B \frac{\partial \mu_{\Delta}}{\partial n_B}\bigg\rvert_{x_p,T} \ ,
\end{align}
often only considering the zero temperature case\cite{Yakovlev:2018jia,Alford:2010gw,Haensel:2000vz,Haensel:1992zz,Sawyer:1989dp}.
As discussed in section \ref{sec:bulk-visc-basics}, because thermal equilibration is so slow in neutrino-transparent nuclear matter in merger conditions, we must use the adiabatic susceptibilities
\begin{align}
 B &= -\frac{1}{n_B} \frac{\partial\mu_{\Delta}}{\partial x_p} \bigg\rvert_{n_B,s/n_B} \ , \\
 C &= n_B \frac{\partial \mu_{\Delta}}{\partial n_B}\bigg\rvert_{x_p,s/n_B} \ .
\end{align}
Note that at zero temperature, adiabatic and isothermal quantities become equivalent\cite{schroeder1999introduction,swendsen2012introduction}.  
Often, it is convenient to work with thermodynamic derivatives at constant temperature $T$, or baryon density $n_B$, or proton fraction $x_p$.  In particular, these three variables are the degrees of freedom in the CompOSE database of EoSs \cite{compose_manual}.  Using a Jacobian coordinate transformation \cite{swendsen2012introduction}, we can relate adiabatic derivatives (derivatives at constant entropy per baryon) to isothermal derivatives.  The adiabatic susceptibility derivatives are related to the isothermal susceptibility
derivatives by
\begin{align}
    \frac{\partial \mu_{\Delta}}{\partial n_B}\bigg \vert_{s/n_B,x_p} &= \frac{\partial \mu_{\Delta}}{\partial n_B}\bigg \vert_{T,x_p} - \frac{\frac{\partial (s/n_B)}{\partial n_B}\big \vert_{T,x_p}\frac{\partial \mu_{\Delta}}{\partial T}\big \vert_{n_B,x_p}}{\frac{\partial (s/n_B)}{\partial T}\big \vert_{n_B,x_p}}\\
    \frac{\partial \mu_{\Delta}}{\partial x_p}\bigg \vert_{s/n_B,n_B} &= \frac{\partial \mu_{\Delta}}{\partial x_p}\bigg \vert_{T,n_B} - \frac{\frac{\partial (s/n_B)}{\partial x_p}\big \vert_{T,n_B}\frac{\partial \mu_{\Delta}}{\partial T}\big \vert_{n_B,x_p}}{\frac{\partial (s/n_B)}{\partial T}\big \vert_{n_B,x_p}}.
\end{align}

The isothermal compressibility is 
\begin{equation}
    \kappa_T^{-1} = n_B \frac{\partial P}{\partial n_B}\bigg\vert_{x_p,T}
\end{equation}
and the adiabatic compressibility is given by
\begin{equation}
    \kappa_S^{-1} = n_B \frac{\partial P}{\partial n_B}\bigg\vert_{x_p,s/n_B}.
\end{equation}
The adiabatic derivative can be obtained from the isothermal derivative by
\begin{equation}
     \frac{\partial P}{\partial n_B}\bigg \vert_{s/n_B,x_p} = \frac{\partial P}{\partial n_B}\bigg \vert_{T,x_p} - \frac{\frac{\partial (s/n_B)}{\partial n_B}\big \vert_{T,x_p}\frac{\partial P}{\partial T}\big \vert_{n_B,x_p}}{\frac{\partial (s/n_B)}{\partial T}\big \vert_{n_B,x_p}}.
\end{equation}

Above $\nsat$, the adiabatic and isothermal derivatives are within 25\% of each other for all temperatures considered here. For nuclear matter that is below $\nsat$ with $T> 5\,\text{MeV}$, there are noticeable differences between the isothermal and adiabatic suseptibility $C$ and the compressibility $\kappa$.  The susceptibility $B$ is not sensitive to differences between adiabaticity and isothermality (the differences are below 10\%).    

Below $\nsat$, the adiabatic $C$ is greater than the isothermal $C$ by as much as a factor of 2.5 (DD2) or 5.5 (IUF).  These large differences are at temperatures above 5\,MeV.  Thus, the adiabatic $C^2/B$ is larger than the isothermal version by factors of up to 6 (DD2) or 30 (IUF).  However, these large differences occur at low densities ($\approx 0.5\nsat$) and high temperatures ($T\approx 10\,\text{MeV}$) where the bulk viscosity is small anyway, since the equilibration rate $\gamma$ is much faster than a 1 kHz density oscillation.  In the regions where bulk viscosity is large, the difference between adiabatic and isothermal susceptibilities is at most a factor of 2 in the quantity $C^2/B$.

At the densities and temperatures where bulk viscosity is large, the isothermal compressibility is at most 20\% larger than the adiabatic compressibility, which means that adiabatic density oscillations would lose energy slightly more slowly than isothermal density oscillations.  At densities below $\nsat$ and temperatures above 5\,MeV, the isothermal compressibility can be up to 40\% (DD2) or 80\% (IUF) larger than the adiabatic value, but the bulk viscosity is too small for fluid elements under these conditions for this to matter.

\bibliographystyle{JHEP}
\bibliography{bulk_visc}{}

\providecommand{\href}[2]{#2}\begingroup\raggedright\begin{thebibliography}{10}

\bibitem{TheLIGOScientific:2017qsa}
{\scshape LIGO Scientific, Virgo} collaboration, B.~P. Abbott et~al.,
  \emph{{GW170817: Observation of Gravitational Waves from a Binary Neutron
  Star Inspiral}},
  \href{http://dx.doi.org/10.1103/PhysRevLett.119.161101}{\emph{Phys. Rev.
  Lett.} {\bf 119} (2017) 161101},
  [\href{https://arxiv.org/abs/1710.05832}{{\tt 1710.05832}}].

\bibitem{Abbott:2018exr}
{\scshape LIGO Scientific, Virgo} collaboration, B.~P. Abbott et~al.,
  \emph{{GW170817: Measurements of neutron star radii and equation of state}},
  \href{http://dx.doi.org/10.1103/PhysRevLett.121.161101}{\emph{Phys. Rev.
  Lett.} {\bf 121} (2018) 161101},
  [\href{https://arxiv.org/abs/1805.11581}{{\tt 1805.11581}}].

\bibitem{Landry:2018prl}
P.~Landry and R.~Essick, \emph{{Nonparametric inference of the neutron star
  equation of state from gravitational wave observations}},
  \href{http://dx.doi.org/10.1103/PhysRevD.99.084049}{\emph{Phys. Rev.} {\bf
  D99} (2019) 084049}, [\href{https://arxiv.org/abs/1811.12529}{{\tt
  1811.12529}}].

\bibitem{Perego:2019adq}
A.~Perego, S.~Bernuzzi and D.~Radice, \emph{{Thermodynamics conditions of
  matter in neutron star mergers}},
  \href{https://arxiv.org/abs/1903.07898}{{\tt 1903.07898}}.

\bibitem{Hanauske:2019qgs}
M.~Hanauske, J.~Steinheimer, A.~Motornenko, V.~Vovchenko, L.~Bovard, E.~R. Most
  et~al., \emph{{Neutron Star Mergers: Probing the EoS of Hot, Dense Matter by
  Gravitational Waves}},
  \href{http://dx.doi.org/10.3390/particles2010004}{\emph{Particles} {\bf 2}
  (2019) 44--56}.

\bibitem{Hanauske:2017oxo}
M.~Hanauske, J.~Steinheimer, L.~Bovard, A.~Mukherjee, S.~Schramm, K.~Takami
  et~al., \emph{{Concluding Remarks: Connecting Relativistic Heavy Ion
  Collisions and Neutron Star Mergers by the Equation of State of Dense Hadron-
  and Quark Matter as signalled by Gravitational Waves}},
  \href{http://dx.doi.org/10.1088/1742-6596/878/1/012031}{\emph{J. Phys. Conf.
  Ser.} {\bf 878} (2017) 012031}.

\bibitem{Baiotti:2016qnr}
L.~Baiotti and L.~Rezzolla, \emph{{Binary neutron star mergers: a review of
  Einstein’s richest laboratory}},
  \href{http://dx.doi.org/10.1088/1361-6633/aa67bb}{\emph{Rept. Prog. Phys.}
  {\bf 80} (2017) 096901}, [\href{https://arxiv.org/abs/1607.03540}{{\tt
  1607.03540}}].

\bibitem{Kastaun:2016elu}
W.~Kastaun, R.~Ciolfi, A.~Endrizzi and B.~Giacomazzo, \emph{{Structure of
  Stable Binary Neutron Star Merger Remnants: Role of Initial Spin}},
  \href{http://dx.doi.org/10.1103/PhysRevD.96.043019}{\emph{Phys. Rev.} {\bf
  D96} (2017) 043019}, [\href{https://arxiv.org/abs/1612.03671}{{\tt
  1612.03671}}].

\bibitem{Bernuzzi:2015opx}
S.~Bernuzzi, D.~Radice, C.~D. Ott, L.~F. Roberts, P.~Moesta and F.~Galeazzi,
  \emph{{How loud are neutron star mergers?}},
  \href{http://dx.doi.org/10.1103/PhysRevD.94.024023}{\emph{Phys. Rev.} {\bf
  D94} (2016) 024023}, [\href{https://arxiv.org/abs/1512.06397}{{\tt
  1512.06397}}].

\bibitem{Foucart:2015gaa}
F.~Foucart, R.~Haas, M.~D. Duez, E.~O'Connor, C.~D. Ott, L.~Roberts et~al.,
  \emph{{Low mass binary neutron star mergers : gravitational waves and
  neutrino emission}},
  \href{http://dx.doi.org/10.1103/PhysRevD.93.044019}{\emph{Phys. Rev.} {\bf
  D93} (2016) 044019}, [\href{https://arxiv.org/abs/1510.06398}{{\tt
  1510.06398}}].

\bibitem{kiuchi:2012mk}
K.~Kiuchi, Y.~Sekiguchi, K.~Kyutoku and M.~Shibata, \emph{{Gravitational waves,
  neutrino emissions, and effects of hyperons in binary neutron star mergers}},
  \href{http://dx.doi.org/10.1088/0264-9381/29/12/124003}{\emph{Class. Quant.
  Grav.} {\bf 29} (2012) 124003}, [\href{https://arxiv.org/abs/1206.0509}{{\tt
  1206.0509}}].

\bibitem{sekiguchi:2011zd}
Y.~Sekiguchi, K.~Kiuchi, K.~Kyutoku and M.~Shibata, \emph{{Gravitational waves
  and neutrino emission from the merger of binary neutron stars}},
  \href{http://dx.doi.org/10.1103/PhysRevLett.107.051102}{\emph{Phys. Rev.
  Lett.} {\bf 107} (2011) 051102}, [\href{https://arxiv.org/abs/1105.2125}{{\tt
  1105.2125}}].

\bibitem{Alford:2017rxf}
M.~G. Alford, L.~Bovard, M.~Hanauske, L.~Rezzolla and K.~Schwenzer,
  \emph{{Viscous Dissipation and Heat Conduction in Binary Neutron-Star
  Mergers}},
  \href{http://dx.doi.org/10.1103/PhysRevLett.120.041101}{\emph{Phys. Rev.
  Lett.} {\bf 120} (2018) 041101},
  [\href{https://arxiv.org/abs/1707.09475}{{\tt 1707.09475}}].

\bibitem{Alford:2018lhf}
M.~G. Alford and S.~P. Harris, \emph{{Beta equilibrium in neutron star
  mergers}}, \href{http://dx.doi.org/10.1103/PhysRevC.98.065806}{\emph{Phys.
  Rev.} {\bf C98} (2018) 065806}, [\href{https://arxiv.org/abs/1803.00662}{{\tt
  1803.00662}}].

\bibitem{Roberts:2016mwj}
L.~F. Roberts and S.~Reddy, \emph{{Charged current neutrino interactions in hot
  and dense matter}},
  \href{http://dx.doi.org/10.1103/PhysRevC.95.045807}{\emph{Phys. Rev.} {\bf
  C95} (2017) 045807}, [\href{https://arxiv.org/abs/1612.02764}{{\tt
  1612.02764}}].

\bibitem{Haensel:1987zz}
P.~Haensel and A.~J. Jerzak, \emph{{Mean free paths of non-degenerate neutrinos
  in neutron star matter}}, {\emph{Astron. Astrophys.} {\bf 179} (1987)
  127--133}.

\bibitem{1979ApJ...230..859S}
R.~F. {Sawyer} and A.~{Soni}, \emph{{Transport of neutrinos in hot neutron-star
  matter}}, \href{http://dx.doi.org/10.1086/157146}{\emph{\apj} {\bf 230}
  (June, 1979) 859--869}.

\bibitem{Sawyer:1975js}
R.~F. Sawyer, \emph{{Neutrino Opacity of Neutron Star Matter}},
  \href{http://dx.doi.org/10.1103/PhysRevD.11.2740}{\emph{Phys. Rev.} {\bf D11}
  (1975) 2740}.

\bibitem{Yakovlev:2000jp}
D.~G. Yakovlev, A.~D. Kaminker, O.~Y. Gnedin and P.~Haensel, \emph{{Neutrino
  emission from neutron stars}},
  \href{http://dx.doi.org/10.1016/S0370-1573(00)00131-9}{\emph{Phys. Rept.}
  {\bf 354} (2001) 1}, [\href{https://arxiv.org/abs/astro-ph/0012122}{{\tt
  astro-ph/0012122}}].

\bibitem{Shapiro:1983du}
S.~L. Shapiro and S.~A. Teukolsky, \emph{{Black holes, white dwarfs, and
  neutron stars: The physics of compact objects}}.
\newblock 1983.

\bibitem{supp}
See Supplemental Material at
  \url{https://journals.aps.org/prc/supplemental/10.1103/PhysRevC.100.035803}
  for tabulated data presented in our figures.

\bibitem{Haensel:2001mw}
P.~Haensel, K.~P. Levenfish and D.~G. Yakovlev, \emph{{Bulk viscosity in
  superfluid neutron star cores. 2. Modified Urca processes in npe mu matter}},
  \href{http://dx.doi.org/10.1051/0004-6361:20010383}{\emph{Astron. Astrophys.}
  {\bf 327} (2001) 130--137},
  [\href{https://arxiv.org/abs/astro-ph/0103290}{{\tt astro-ph/0103290}}].

\bibitem{Yuan:2005tj}
Y.-F. Yuan, \emph{{Electron positron capture rates and the steady state
  equilibrium condition for electron-positron plasma with nucleons}},
  \href{http://dx.doi.org/10.1103/PhysRevD.72.013007}{\emph{Phys. Rev.} {\bf
  D72} (2005) 013007}, [\href{https://arxiv.org/abs/astro-ph/0507311}{{\tt
  astro-ph/0507311}}].

\bibitem{Haensel:2000vz}
P.~Haensel, K.~P. Levenfish and D.~G. Yakovlev, \emph{{Bulk viscosity in
  superfluid neutron star cores. I. direct urca processes in npe mu matter}},
  {\emph{Astron. Astrophys.} {\bf 357} (2000) 1157--1169},
  [\href{https://arxiv.org/abs/astro-ph/0004183}{{\tt astro-ph/0004183}}].

\bibitem{1992A&A...262..131H}
P.~{Haensel}, \emph{{Non-equilibrium neutrino emissivities and opacities of
  neutron star matter}}, {\emph{Astron. Astrophys.} {\bf 262} (Aug., 1992)
  131--137}.

\bibitem{fi09000a}
A.~Finzi and R.~A. Wolf, \emph{Hot, vibrating neutron stars},
  \href{http://dx.doi.org/10.1086/149708}{\emph{Astrophys. J.} {\bf 153} (1968)
  835--848}.

\bibitem{Alford:2010gw}
M.~G. Alford, S.~Mahmoodifar and K.~Schwenzer, \emph{{Large amplitude behavior
  of the bulk viscosity of dense matter}},
  \href{http://dx.doi.org/10.1088/0954-3899/37/12/125202}{\emph{J. Phys.} {\bf
  G37} (2010) 125202}, [\href{https://arxiv.org/abs/1005.3769}{{\tt
  1005.3769}}].

\bibitem{Schmitt:2017efp}
A.~Schmitt and P.~Shternin, \emph{{Reaction rates and transport in neutron
  stars}},
  \href{http://dx.doi.org/10.1007/978-3-319-97616-7_9}{\emph{Astrophys. Space
  Sci. Libr.} {\bf 457} (2018) 455--574},
  [\href{https://arxiv.org/abs/1711.06520}{{\tt 1711.06520}}].

\bibitem{Haensel:1992zz}
P.~Haensel and R.~Schaeffer, \emph{{Bulk viscosity of hot-neutron-star matter
  from direct URCA processes}},
  \href{http://dx.doi.org/10.1103/PhysRevD.45.4708}{\emph{Phys. Rev.} {\bf D45}
  (1992) 4708--4712}.

\bibitem{Reisenegger:1994be}
A.~Reisenegger, \emph{{Deviations from chemical equilibrium due to spindown as
  an internal heat source in neutron stars}},
  \href{http://dx.doi.org/10.1086/175480}{\emph{Astrophys. J.} {\bf 442} (1995)
  749}, [\href{https://arxiv.org/abs/astro-ph/9410035}{{\tt
  astro-ph/9410035}}].

\bibitem{Friman:1978zq}
B.~L. Friman and O.~V. Maxwell, \emph{{Neutron Star Neutrino Emissivities}},
  \href{http://dx.doi.org/10.1086/157313}{\emph{Astrophys. J.} {\bf 232} (1979)
  541--557}.

\bibitem{Sawyer:1989dp}
R.~F. Sawyer, \emph{{Bulk viscosity of hot neutron-star matter and the maximum
  rotation rates of neutron stars}},
  \href{http://dx.doi.org/10.1103/PhysRevD.39.3804}{\emph{Phys. Rev.} {\bf D39}
  (1989) 3804--3806}.

\bibitem{Hempel:2009mc}
M.~Hempel and J.~Schaffner-Bielich, \emph{{Statistical Model for a Complete
  Supernova Equation of State}},
  \href{http://dx.doi.org/10.1016/j.nuclphysa.2010.02.010}{\emph{Nucl. Phys.}
  {\bf A837} (2010) 210--254}, [\href{https://arxiv.org/abs/0911.4073}{{\tt
  0911.4073}}].

\bibitem{Typel:2009sy}
S.~Typel, G.~Ropke, T.~Klahn, D.~Blaschke and H.~H. Wolter, \emph{{Composition
  and thermodynamics of nuclear matter with light clusters}},
  \href{http://dx.doi.org/10.1103/PhysRevC.81.015803}{\emph{Phys. Rev.} {\bf
  C81} (2010) 015803}, [\href{https://arxiv.org/abs/0908.2344}{{\tt
  0908.2344}}].

\bibitem{Fattoyev:2010mx}
F.~J. Fattoyev, C.~J. Horowitz, J.~Piekarewicz and G.~Shen, \emph{{Relativistic
  effective interaction for nuclei, giant resonances, and neutron stars}},
  \href{http://dx.doi.org/10.1103/PhysRevC.82.055803}{\emph{Phys. Rev.} {\bf
  C82} (2010) 055803}, [\href{https://arxiv.org/abs/1008.3030}{{\tt
  1008.3030}}].

\bibitem{RocaMaza:2008ja}
X.~Roca-Maza and J.~Piekarewicz, \emph{{Impact of the symmetry energy on the
  outer crust of non-accreting neutron stars}},
  \href{http://dx.doi.org/10.1103/PhysRevC.78.025807}{\emph{Phys. Rev.} {\bf
  C78} (2008) 025807}, [\href{https://arxiv.org/abs/0805.2553}{{\tt
  0805.2553}}].

\bibitem{glendenning2000compact}
N.~Glendenning, \emph{Compact Stars: Nuclear Physics, Particle Physics, and
  General Relativity}.
\newblock Astronomy and Astrophysics Library. Springer New York, 2000.

\bibitem{Dutra:2014qga}
M.~Dutra, O.~Lourenço, S.~S. Avancini, B.~V. Carlson, A.~Delfino, D.~P.
  Menezes et~al., \emph{{Relativistic Mean-Field Hadronic Models under Nuclear
  Matter Constraints}},
  \href{http://dx.doi.org/10.1103/PhysRevC.90.055203}{\emph{Phys. Rev.} {\bf
  C90} (2014) 055203}, [\href{https://arxiv.org/abs/1405.3633}{{\tt
  1405.3633}}].

\bibitem{Oertel:2016bki}
M.~Oertel, M.~Hempel, T.~Klähn and S.~Typel, \emph{{Equations of state for
  supernovae and compact stars}},
  \href{http://dx.doi.org/10.1103/RevModPhys.89.015007}{\emph{Rev. Mod. Phys.}
  {\bf 89} (2017) 015007}, [\href{https://arxiv.org/abs/1610.03361}{{\tt
  1610.03361}}].

\bibitem{Roberts:2012um}
L.~F. Roberts, S.~Reddy and G.~Shen, \emph{{Medium modification of the charged
  current neutrino opacity and its implications}},
  \href{http://dx.doi.org/10.1103/PhysRevC.86.065803}{\emph{Phys. Rev.} {\bf
  C86} (2012) 065803}, [\href{https://arxiv.org/abs/1205.4066}{{\tt
  1205.4066}}].

\bibitem{Fu:2008zzg}
W.-j. Fu, G.-h. Wang and Y.-x. Liu, \emph{{Electron Capture and Its Reverse
  Process in Hot and Dense Astronuclear Matter}},
  \href{http://dx.doi.org/10.1086/528361}{\emph{Astrophys. J.} {\bf 678} (2008)
  1517--1529}.

\bibitem{Leinson:2002bw}
L.~B. Leinson, \emph{{Direct Urca processes on nucleons in cooling neutron
  stars}}, \href{http://dx.doi.org/10.1016/S0375-9474(02)00991-0}{\emph{Nucl.
  Phys.} {\bf A707} (2002) 543--560},
  [\href{https://arxiv.org/abs/hep-ph/0207116}{{\tt hep-ph/0207116}}].

\bibitem{schroeder1999introduction}
D.~Schroeder, \emph{An Introduction to Thermal Physics}.
\newblock Addison Wesley, 1999.

\bibitem{compose_manual}
``Compose reference manual.'' \url{https://compose.obspm.fr/manual/}.

\bibitem{Schmitt:2010pn}
A.~Schmitt, \emph{{Dense matter in compact stars: A pedagogical introduction}},
  \href{http://dx.doi.org/10.1007/978-3-642-12866-0}{\emph{Lect. Notes Phys.}
  {\bf 811} (2010) 1--111}, [\href{https://arxiv.org/abs/1001.3294}{{\tt
  1001.3294}}].

\bibitem{Dexheimer:2007mt}
V.~A. Dexheimer, C.~A.~Z. Vasconcellos and B.~E.~J. Bodmann, \emph{{On the
  Density Dependent Nuclear Matter Compressibility}},
  \href{http://dx.doi.org/10.1103/PhysRevC.77.065803}{\emph{Phys. Rev.} {\bf
  C77} (2008) 065803}, [\href{https://arxiv.org/abs/0708.0131}{{\tt
  0708.0131}}].

\bibitem{Sawyer:1980wp}
R.~F. Sawyer, \emph{{Damping of neutron star pulsations by weak interaction
  processes}}, \href{http://dx.doi.org/10.1086/157858}{\emph{Astrophys. J.}
  {\bf 237} (1980) 187--197}.

\bibitem{1990ApJ...363..603C}
C.~{Cutler}, L.~{Lindblom} and R.~J. {Splinter}, \emph{{Damping times for
  neutron star oscillations}},
  \href{http://dx.doi.org/10.1086/169370}{\emph{\apj} {\bf 363} (Nov., 1990)
  603--611}.

\bibitem{Hotokezaka:2013iia}
K.~Hotokezaka, K.~Kiuchi, K.~Kyutoku, T.~Muranushi, Y.-i. Sekiguchi, M.~Shibata
  et~al., \emph{{Remnant massive neutron stars of binary neutron star mergers:
  Evolution process and gravitational waveform}},
  \href{http://dx.doi.org/10.1103/PhysRevD.88.044026}{\emph{Phys. Rev.} {\bf
  D88} (2013) 044026}, [\href{https://arxiv.org/abs/1307.5888}{{\tt
  1307.5888}}].

\bibitem{Chaurasia:2018zhg}
S.~V. Chaurasia, T.~Dietrich, N.~K. Johnson-McDaniel, M.~Ujevic, W.~Tichy and
  B.~Brügmann, \emph{{Gravitational waves and mass ejecta from binary neutron
  star mergers: Effect of large eccentricities}},
  \href{http://dx.doi.org/10.1103/PhysRevD.98.104005}{\emph{Phys. Rev.} {\bf
  D98} (2018) 104005}, [\href{https://arxiv.org/abs/1807.06857}{{\tt
  1807.06857}}].

\bibitem{Clark:2015zxa}
J.~A. Clark, A.~Bauswein, N.~Stergioulas and D.~Shoemaker, \emph{{Observing
  Gravitational Waves From The Post-Merger Phase Of Binary Neutron Star
  Coalescence}},
  \href{http://dx.doi.org/10.1088/0264-9381/33/8/085003}{\emph{Class. Quant.
  Grav.} {\bf 33} (2016) 085003}, [\href{https://arxiv.org/abs/1509.08522}{{\tt
  1509.08522}}].

\bibitem{Rezzolla:2016nxn}
L.~Rezzolla and K.~Takami, \emph{{Gravitational-wave signal from binary neutron
  stars: a systematic analysis of the spectral properties}},
  \href{http://dx.doi.org/10.1103/PhysRevD.93.124051}{\emph{Phys. Rev.} {\bf
  D93} (2016) 124051}, [\href{https://arxiv.org/abs/1604.00246}{{\tt
  1604.00246}}].

\bibitem{Dietrich:2016hky}
T.~Dietrich, M.~Ujevic, W.~Tichy, S.~Bernuzzi and B.~Bruegmann,
  \emph{{Gravitational waves and mass ejecta from binary neutron star mergers:
  Effect of the mass-ratio}},
  \href{http://dx.doi.org/10.1103/PhysRevD.95.024029}{\emph{Phys. Rev.} {\bf
  D95} (2017) 024029}, [\href{https://arxiv.org/abs/1607.06636}{{\tt
  1607.06636}}].

\bibitem{Maione:2017aux}
F.~Maione, R.~De~Pietri, A.~Feo and F.~Löffler, \emph{{Spectral analysis of
  gravitational waves from binary neutron star merger remnants}},
  \href{http://dx.doi.org/10.1103/PhysRevD.96.063011}{\emph{Phys. Rev.} {\bf
  D96} (2017) 063011}, [\href{https://arxiv.org/abs/1707.03368}{{\tt
  1707.03368}}].

\bibitem{Bauswein:2018bma}
A.~Bauswein, N.-U.~F. Bastian, D.~B. Blaschke, K.~Chatziioannou, J.~A. Clark,
  T.~Fischer et~al., \emph{{Identifying a first-order phase transition in
  neutron star mergers through gravitational waves}},
  \href{http://dx.doi.org/10.1103/PhysRevLett.122.061102}{\emph{Phys. Rev.
  Lett.} {\bf 122} (2019) 061102},
  [\href{https://arxiv.org/abs/1809.01116}{{\tt 1809.01116}}].

\bibitem{Baiotti:2019sew}
L.~Baiotti, \emph{{Gravitational waves from neutron star mergers and their
  relation to the nuclear equation of state}},
  \href{https://arxiv.org/abs/1907.08534}{{\tt 1907.08534}}.

\bibitem{DePietri:2019khb}
R.~De~Pietri, A.~Drago, A.~Feo, G.~Pagliara, M.~Pasquali, S.~Traversi et~al.,
  \emph{{Merger of compact stars in the two-families scenario}},
  \href{https://arxiv.org/abs/1904.01545}{{\tt 1904.01545}}.

\bibitem{Alford:2019kdw}
M.~Alford, A.~Harutyunyan and A.~Sedrakian, \emph{{Bulk viscosity of baryonic
  matter with trapped neutrinos}},
  \href{https://arxiv.org/abs/1907.04192}{{\tt 1907.04192}}.

\bibitem{Ardevol-Pulpillo:2018btx}
R.~Ardevol-Pulpillo, H.~T. Janka, O.~Just and A.~Bauswein, \emph{{Improved
  Leakage-Equilibration-Absorption Scheme (ILEAS) for Neutrino Physics in
  Compact Object Mergers}},
  \href{http://dx.doi.org/10.1093/mnras/stz613}{\emph{Mon. Not. Roy. Astron.
  Soc.} {\bf 485} (2019) 4754--4789},
  [\href{https://arxiv.org/abs/1808.00006}{{\tt 1808.00006}}].

\bibitem{Perego:2014qda}
A.~Perego, E.~Gafton, R.~Cabezón, S.~Rosswog and M.~Liebendörfer,
  \emph{{MODA: a new algorithm to compute optical depths in multidimensional
  hydrodynamic simulations}},
  \href{http://dx.doi.org/10.1051/0004-6361/201423755}{\emph{Astron.
  Astrophys.} {\bf 568} (2014) A11},
  [\href{https://arxiv.org/abs/1403.1297}{{\tt 1403.1297}}].

\bibitem{Galeazzi:2013mia}
F.~Galeazzi, W.~Kastaun, L.~Rezzolla and J.~A. Font, \emph{{Implementation of a
  simplified approach to radiative transfer in general relativity}},
  \href{http://dx.doi.org/10.1103/PhysRevD.88.064009}{\emph{Phys. Rev.} {\bf
  D88} (2013) 064009}, [\href{https://arxiv.org/abs/1306.4953}{{\tt
  1306.4953}}].

\bibitem{Sekiguchi:2012uc}
Y.~Sekiguchi, K.~Kiuchi, K.~Kyutoku and M.~Shibata, \emph{{Current Status of
  Numerical-Relativity Simulations in Kyoto}},
  \href{http://dx.doi.org/10.1093/ptep/pts011}{\emph{PTEP} {\bf 2012} (2012)
  01A304}, [\href{https://arxiv.org/abs/1206.5927}{{\tt 1206.5927}}].

\bibitem{Rosswog:2003rv}
S.~Rosswog and M.~Liebendoerfer, \emph{{High resolution calculations of merging
  neutron stars. 2: Neutrino emission}},
  \href{http://dx.doi.org/10.1046/j.1365-8711.2003.06579.x}{\emph{Mon. Not.
  Roy. Astron. Soc.} {\bf 342} (2003) 673},
  [\href{https://arxiv.org/abs/astro-ph/0302301}{{\tt astro-ph/0302301}}].

\bibitem{1999JCoAM.109..281M}
A.~{Mezzacappa} and O.~E.~B. {Messer}, \emph{{Neutrino transport in core
  collapse supernovae.}}, {\emph{Journal of Computational and Applied
  Mathematics} {\bf 109} (Sept., 1999) 281--319}.

\bibitem{Grill:2014aea}
F.~Grill, H.~Pais, C.~Providência, I.~Vidaña and S.~S. Avancini,
  \emph{{Equation of state and thickness of the inner crust of neutron stars}},
  \href{http://dx.doi.org/10.1103/PhysRevC.90.045803}{\emph{Phys. Rev.} {\bf
  C90} (2014) 045803}, [\href{https://arxiv.org/abs/1404.2753}{{\tt
  1404.2753}}].

\bibitem{Fattoyev:2017zhb}
F.~J. Fattoyev, C.~J. Horowitz and B.~Schuetrumpf, \emph{{Quantum Nuclear Pasta
  and Nuclear Symmetry Energy}},
  \href{http://dx.doi.org/10.1103/PhysRevC.95.055804}{\emph{Phys. Rev.} {\bf
  C95} (2017) 055804}, [\href{https://arxiv.org/abs/1703.01433}{{\tt
  1703.01433}}].

\bibitem{Pais:2015lma}
H.~Pais, C.~Providência, W.~G. Newton and J.~R. Stone, \emph{{Core-collapse
  supernova matter: light clusters, pasta phase and phase transitions}},  in
  \emph{{Proceedings, Compact Stars in the QCD Phase Diagram IV (CSQCD IV):
  Prerow, Germany, September 26-30, 2014}}, 2015.
\newblock \href{https://arxiv.org/abs/1503.08753}{{\tt 1503.08753}}.

\bibitem{Pais:2014hoa}
H.~Pais, W.~G. Newton and J.~R. Stone, \emph{{Phase transitions in
  Core-Collapse Supernova Matter at sub-saturation densities}},
  \href{http://dx.doi.org/10.1103/PhysRevC.90.065802}{\emph{Phys. Rev.} {\bf
  C90} (2014) 065802}, [\href{https://arxiv.org/abs/1411.1885}{{\tt
  1411.1885}}].

\bibitem{Oyamatsu:1993zz}
K.~Oyamatsu, \emph{{Nuclear shapes in the inner crust of a neutron star}},
  \href{http://dx.doi.org/10.1016/0375-9474(93)90020-X}{\emph{Nucl. Phys.} {\bf
  A561} (1993) 431--452}.

\bibitem{Yakovlev:2018jia}
D.~G. Yakovlev, M.~E. Gusakov and P.~Haensel, \emph{{Bulk viscosity in a
  neutron star mantle}},
  \href{http://dx.doi.org/10.1093/mnras/sty2639}{\emph{Mon. Not. Roy. Astron.
  Soc.} {\bf 481} (2018) 4924}, [\href{https://arxiv.org/abs/1809.08609}{{\tt
  1809.08609}}].

\bibitem{Gusakov:2004mj}
M.~E. Gusakov, D.~G. Yakovlev, P.~Haensel and O.~Y. Gnedin, \emph{{Direct Urca
  process in a neutron star mantle}},
  \href{http://dx.doi.org/10.1051/0004-6361:20040288}{\emph{Astron. Astrophys.}
  {\bf 421} (2004) 1143--1148},
  [\href{https://arxiv.org/abs/astro-ph/0404165}{{\tt astro-ph/0404165}}].

\bibitem{Roggero:2017pag}
A.~Roggero, J.~Margueron, L.~F. Roberts and S.~Reddy, \emph{{Nuclear pasta in
  hot dense matter and its implications for neutrino scattering}},
  \href{http://dx.doi.org/10.1103/PhysRevC.97.045804}{\emph{Phys. Rev.} {\bf
  C97} (2018) 045804}, [\href{https://arxiv.org/abs/1710.10206}{{\tt
  1710.10206}}].

\bibitem{Alford:2010jf}
M.~G. Alford and G.~Good, \emph{{Leptonic contribution to the bulk viscosity of
  nuclear matter}},
  \href{http://dx.doi.org/10.1103/PhysRevC.82.055805}{\emph{Phys. Rev.} {\bf
  C82} (2010) 055805}, [\href{https://arxiv.org/abs/1003.1093}{{\tt
  1003.1093}}].

\bibitem{Shternin:2018dcn}
P.~S. Shternin, M.~Baldo and P.~Haensel, \emph{{In-medium enhancement of the
  modified Urca neutrino reaction rates}},
  \href{http://dx.doi.org/10.1016/j.physletb.2018.09.035}{\emph{Phys. Lett.}
  {\bf B786} (2018) 28--34}, [\href{https://arxiv.org/abs/1807.06569}{{\tt
  1807.06569}}].

\bibitem{swendsen2012introduction}
R.~Swendsen, \emph{An Introduction to Statistical Mechanics and
  Thermodynamics}.
\newblock Oxford Graduate Texts. OUP Oxford, 2012.

\end{thebibliography}\endgroup

\end{document}